\documentclass[acmsmall]{acmart}

\usepackage{booktabs}
\usepackage[skip=0pt]{caption}
\usepackage{subcaption}
\usepackage{multirow}
\usepackage[table]{colortbl}
\usepackage[inline]{enumitem}
\usepackage{lscape}
\usepackage{balance}
\usepackage{footnote}
\usepackage{rotating}
\makesavenoteenv{tabular}
\makesavenoteenv{table}

\acmJournal{TOIS}
\acmVolume{39}
\acmNumber{4}
\acmArticle{111}
\acmMonth{4}

\copyrightyear{2021}
\acmYear{2021}
\setcopyright{acmlicensed}

\allowdisplaybreaks

\author{Svitlana Vakulenko}
\affiliation{%
  \institution{University of Amsterdam}
  \city{Amsterdam} 
  \country{The Netherlands}}
\email{s.vakulenko@uva.nl}

\author{Evangelos Kanoulas}
\affiliation{%
  \institution{University of Amsterdam}
  \city{Amsterdam} 
  \country{The Netherlands}}
\email{e.kanoulas@uva.nl}

\author{Maarten de Rijke}
\affiliation{%
  \institution{University of Amsterdam \& Ahold Delhaize}
  \city{Amsterdam} 
  \country{The Netherlands}}
\email{m.derijke@uva.nl}

\usepackage[export]{adjustbox}

\usepackage{amsthm}

\theoremstyle{definition}
\newtheorem{definition}{Definition}[section]
\newcommand{\revised}[1]{\textcolor{black}{#1}}

\begin{document}

% Dove va Conversational Search?
% \title{Discovering Measurable Dimensions of Mixed Initiative in Information-Seeking Dialogues}
\title{A Large-Scale Analysis of Mixed Initiative in Information-Seeking Dialogues for Conversational Search}
% \title{An Extensive Analysis of Mixed Initiative and Collaboration in Information-Seeking Dialogues}
%https://www.overleaf.com/project/5f44ece9ca77da0001a866fc

\renewcommand{\shortauthors}{Vakulenko, Kanoulas and de Rijke}

\begin{abstract}
  Conversational search is a relatively young area of research that aims at automating an information-seeking dialogue. In this paper we help to position it with respect to other research areas within conversational Artificial Intelligence (AI) by analysing the structural properties of an information-seeking dialogue. To this end, we perform a large-scale dialogue analysis of more than 150K transcripts from 16 publicly available dialogue datasets. These datasets were collected to inform different dialogue-based tasks including conversational search.
  We extract different patterns of mixed initiative from these dialogue transcripts and use them to compare dialogues of different types.
Moreover, we contrast the patterns found in information-seeking dialogues that are being used for research purposes with the patterns found in virtual reference interviews that were conducted by professional librarians.
The insights we provide \begin{enumerate*}\item establish close relations between conversational search and other conversational AI tasks; and \item uncover limitations of existing conversational datasets to inform future data collection tasks.\end{enumerate*}
\end{abstract}




\setcopyright{acmlicensed}
\acmJournal{TOIS}
\acmYear{2021} \acmVolume{1} \acmNumber{1} \acmArticle{1} \acmMonth{1} \acmPrice{15.00}\acmDOI{10.1145/3466796}

\maketitle

\section{Introduction}
\label{sec:introduction}

\begin{figure}[t]
  \centering
  \includegraphics[width=\linewidth]{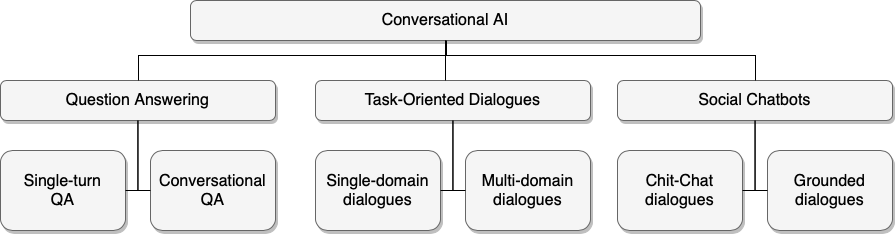}
  \caption{Tasks within conversational AI as identified by~\citet{INR-074}. Where do conversational search and information-seeking dialogues fit in this picture?}
  \label{fig:survey}
\end{figure}

% introduce conversational AI
Research in conversational AI spans across multiple disciplines, including natural language processing, information retrieval, machine learning and dialogue systems.
Its main goal is the development of intelligent conversational systems, which find application across a wide range of domains, such as customer support, education, e-commerce, health, entertainment etc.
% define conversational AI in terms of its subtasks
Several subtasks have been proposed in the context of conversational AI.
A recent survey of state-of-art systems in conversational AI groups them according to three main tasks: question answering, task-oriented dialogues, and social chatbots~\citep{INR-074}; see Figure~\ref{fig:survey}.

% introduce conversational search
Conversational search has been recognised as an important new frontier within information retrieval~\citep{anand_et_al:DR:2020:11983}.
But how does conversational search relate to the tasks previously proposed in the context of conversational AI?
% define conversational search
Several definitions of conversational search have been proposed to date~\cite{anand_et_al:DR:2020:11983}.
All of them agree that the task of conversational search is the development of systems that enable information retrieval by means of a natural-language interaction, i.e., dialogue interface~\cite{radlinski2017theoretical,trippas2020towards,svitlana-vakulenko-phd-thesis-2019}.
This type of dialogue is referred to as an \emph{information-seeking dialogue}.
In this way, conversational search is defined in terms of the type of dialogue it is expected to produce.

% introduce information
Since conversational search is aimed at developing systems that are able to support an information-seeking dialogue, we need to understand the nature and properties of this type of dialogue so as to be able to determine its success.
What is an information-seeking dialogue?
What are the main characteristics that differentiate it from other dialogue types?
Our main goal in this paper is to leverage conversational data for conducting an empirical analysis of information-seeking dialogues collected to date.
We aim to gain a better understanding of how information-seeking dialogues differ from dialogues that were collected in the context of other conversational AI tasks, such as chit-chat, conversational QA, and task-oriented dialogue.

In the early days of information retrieval research, reference interviews with a librarian formed an important source of inspiration and information~\cite{taylor1968,DBLP:conf/riao/DanielsBB85,belkin1995cases}.
For the purpose of our study, we collaborated with a global library cooperative (OCLC) to obtain a sample of 560 anonymized virtual reference interviews conducted on-line by professional librarians in 2010.
To the best of our knowledge, this is the first systematic study of this scale designed to compare the patterns of interactions observed in virtual reference interviews with the patterns extracted from publicly available dialogue datasets.
Our goal was to identify datasets that can best represent the type of interaction that is characteristic of a professional reference interview.
The results of such an analysis lead to a recommendation for the datasets that can be used for training and evaluation of conversational search systems designed to mimic this dialogue type.

% RQs
Our main research questions can be summarised as follows:
\begin{enumerate}[label=(RQ\arabic*),leftmargin=*]
    \item What are the structural properties of information-seeking dialogues that differentiate them from other dialogue types?
    \item Which datasets contain dialogues similar to virtual reference interviews conducted by professional librarians?
\end{enumerate}

% contributions
% By answering these questions we aim to provide guidelines on 
% \begin{enumerate*}
% \item which datasets should be used for developing conversational search systems; and \item what the output of a conversational search system should look like.
% \end{enumerate*}

% our means: patterns of mixed initiative
\noindent%
To answer our research questions we introduce ConversationShape, a framework for dialogue analysis.
We focus on the patterns of mixed initiative since the asymmetry of roles was previously identified as the innate property of an information-seeking dialogue, which is due to the knowledge distribution between the conversation participants~\cite{birrerissa}.
Mixed initiative was also identified as one of the core requirements for a conversational search system~\cite{radlinski2017theoretical}.
Therefore, identifying mixed initiative and describing its use in a successful information-seeking conversation is important for the design and evaluation of conversational search systems.

% our contributions
We demonstrate how our dialogue analysis framework can be applied in practice to conduct a large-scale analysis covering all available conversational datasets.
The results of our analysis help us to define an information-seeking dialogue in terms of a set of measurable properties, and to demonstrate their similarities with, and differences to, other dialogue types.

Our contributions include
\begin{enumerate*}
\item a large-scale analysis characterising the mix of initiative across different dialogue tasks; and
\item ConversationShape, a methodological framework that has been used to conduct this dialogue analysis and can be re-used to position new dialogue datasets with respect to existing ones.
\end{enumerate*}

% paper structure
The remainder of the paper is organised as follows.
Section~\ref{sec:related_work} sets the background by providing an overview of existing theoretical foundations for conversational search, previous research focused on dialogue analysis, and mixed-initiative systems.
We also review previous studies that manually analyse dialogue transcripts to formulate grounded theories of information-seeking dialogues and mixed initiative in dialogue.
\revised{We then proceed to describe the ConversationShape analysis framework in Section~\ref{sec:framework}, consisting of fingerprinting, dialogue flow, and asymmetry metrics. Section~\ref{sec:experimental_setup} details our experimental setup and introduces the dialogue datasets that we consider in our analysis.
We report the results of applying the ConversationShape framework to these datasets in Section~\ref{sec:results}, reflect on the outcomes in Section~\ref{sec:discussion}, and share our conclusions in Section~\ref{sec:consclusion}}.

\section{Related Work}
\label{sec:related_work}

% how does our work relate to previous work on which dimensions?
Our work contributes to the body of research devoted to analysing information-seeking dialogues.
This type of analysis is useful for informing the theoretical foundations of conversational search by grounding it in empirical observations.
Such observations can then be used directly to propose new dialogue models~\cite{trippas2018informing,DBLP:conf/ecir/VakulenkoRCR19}.

% section structure
We start by briefly summarising existing theories of conversational search and information-seeking dialogues. The second subsection provides an overview of the previous research that studies transcripts of information-seeking dialogues and the approaches used to analyse their discourse structure. In the last subsection we review the work focusing specifically on the mix of initiative in dialogues of different types.

\subsection{Theories of Conversational Search}

% theories need grounding in empirical data
Several theories of conversational search have been proposed to date~\cite{radlinski2017theoretical,azzopardi2018conceptualizing,trippas2020towards}.
They are mainly concerned with modeling the set of interactions that occur in the context of an information-seeking dialogue.
\revised{These ideas can be traced back to the Conversational Roles (COR) Model for generating information-seeking dialogues~\cite{DBLP:journals/ipm/SitterS92}.}

% \todo{expand with details on theories and models}
\citet{radlinski2017theoretical} proposed a conversational search model, in which the system interactively provides information to the user and incorporates feedback to elicit user preferences.
The goal of the system is to maximise user satisfaction by finding the items with maximal utility.
\citet{azzopardi2018conceptualizing} proposed an extended set of user-system actions and subtasks that include suggestion, explanation, navigation, interruption and interrogation.
% It is important to note that both of the aforementioned theories are not based on any systematic analysis of empirical data.

An important limitation of these theoretical models describing an information-seeking behaviour, in general, is that they are often based on anecdotal evidence drawn from a handful of dialogue transcripts rather than on systematic empirical observations.
In contrast to prior work, \citet{trippas2020towards} performed thematic analysis of the information-seeking dialogue datasets (SCSdata and MISC) by manually labeling 1,710 utterances from 53 dialogue transcripts.
Such analysis allowed them to develop a fine-grained labeling schema, SCoSAS, describing the interactions in conversational search.
SCoSAS includes 84 labels grouped into three main themes: Task Level, Discourse Lever and Other.
Unfortunately, this approach for grounded theory building through dialogue analysis does not scale since it relies on manual annotations of dialogue transcripts.
Training a supervised classifier requires a large number of annotated samples, which are not available on such a fine-grained level. 
% relation to us
Our goal in this paper is to establish a mechanism that can enable us to continuously refine the theories of conversational search based on the growing volume of empirical data.

% mixed initiative is yet poorly understood
There are major gaps in our understanding of the specifics of dialogue interactions that occur in the context of conversational search~\cite{thomas2020theories}.
In particular, \citet{radlinski2017theoretical} contribute a fundamental set of requirements for a conversational search system, which includes \emph{mixed initiative}, \emph{user revealment} and \emph{system revealment}.
The mechanisms underlying these concepts are yet poorly understood.
% relation to us
To fill this gap, we conduct a large-scale dialogue analysis that focuses on several dimensions of mixed initiative in dialogues.

\begin{figure}[tbh!]
  \centering
  \includegraphics[clip,trim=10mm 0mm 0mm 0mm,width=0.7\linewidth]{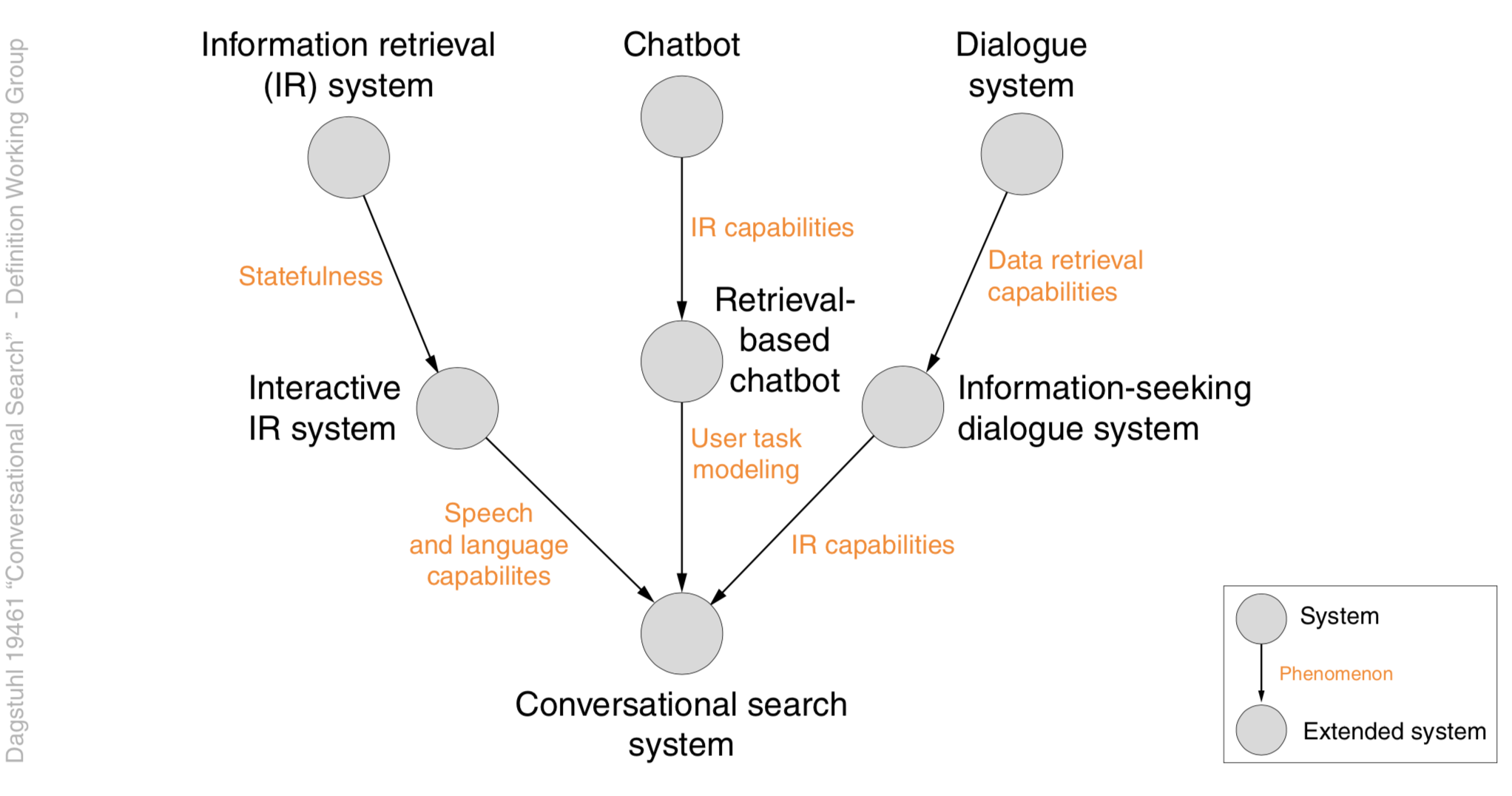}
  \caption{The Dagstuhl Typology of Conversational Search~\cite{anand_et_al:DR:2020:11983} describes how existing systems can evolve into a conversational search system by dropping different assumptions or limitations.}
  \label{fig:dtcs}
\end{figure}

It is important to position conversational search with respect to other research tasks and disciplines.
The first Dagstuhl seminar on Conversational Search took an important step in this direction by introducing a typology of conversational search systems~\cite{anand_et_al:DR:2020:11983}; see Figure~\ref{fig:dtcs}.
This typology relates conversational search systems to other types of interactive system, such as chatbots, information retrieval and dialogue systems.
In essence, it describes how existing systems can evolve into a conversational search system by gradually extending their capabilities.

The Dagstuhl typology provides an alignment based on the design of existing systems.
This perspective has its limitations in biasing the design of a new system towards previously proposed approaches.
% relation to us
We take a fundamentally different approach by looking at the expected output of the systems irrespective of their architectural design. 
To understand the difference between systems we analyse the differences between the dialogues that these systems are designed to produce.

\subsection{Dialogue Analysis}

Since the 1980s discourse analysis has been applied to characterise the content of information-seeking dialogues.
\citet{belkin1982ask} suggest that dialogue analysis should be applied to collect patterns from human-human dialogues and identify different information-seeking strategies.
These patterns can then serve as reusable scripts within an interactive information retrieval system.

\citet{DBLP:conf/riao/DanielsBB85} collect six transcripts of online reference interviews. They use one of the dialogues to derive a set of goals that both intermediary and seeker pursue in a conversation, and use them to annotate utterances in the remaining five dialogue transcripts. Their set of goals for document retrieval interactions include 23 labels grouped into 8 foci, such as Problem Description, Retrieval Strategies, Response Generator, Explanation, etc.
In later work, \citet{saracevic1997users} perform a similar type of analysis on 40 transcripts of video recordings of reference interviews. They derive their own classification schema that distinguishes utterances across 8 different categories, such as System explanation, Back channeling, Search tactic, etc.

% lab studies spoken video recording non-professional intermediaries
In more recent work, \citet{trippas2018informing} reproduce this approach when collecting the Spoken Conversational Search (SCSData) dataset.
However, instead of recording interactions with trained intermediaries, such as professional librarians, they recruit volunteers (mostly students) to play the roles of seekers and intermediaries.
A similar approach has been employed to collect the MISC corpus~\citep{thomas2017misc}.
The authors use thematic analysis to derive a new annotation schema (SCoSAS) and annotate both datasets (SCSData and MISC) using this schema~\cite{trippas2020towards}.
They note important differences in the interaction patterns discovered between the two datasets. These differences are attributed to the difference in task formulation and errors from automated speech recognition.

% tech support chat/forum qa forum
Next, motivated by the lack of data for training machine learning models, \citet{InforSeek_Response_Ranking} collect forum threads of on-line technical support platforms as samples of information-seeking dialogues.
This approach allows the authors to create the large-scale MSDialog corpus with 35.5K dialogues.
The authors reuse an existing taxonomy for classifying forum messages with user intents and extend it with four new labels.
This taxonomy has been used to annotate a subset of the MSDialog dataset (2K dialogues) using crowdsourcing.
Then, the authors extract sequences of utterance labels as dialogue flow patterns and compare them with patterns extracted from the Ubuntu Dialog Corpus~\citep{DBLP:conf/sigdial/LowePSP15}. The authors report that while the extracted patterns are the same for both dialogue datasets, their relative frequencies are different.

% similarity between all these studies
All of the dialogue analysis studies mentioned above follow the same procedure: (1)~derive a set of labels from a sample of the conversational data; (2)~annotate utterances with these labels; and (3)~extract patterns using these labels, such as label co-occurrences (n-grams).
% evaluate: what is positive/negative about this approach
The benefit of this approach is that it allows us to perform different levels of analysis using different annotation schemas.
The obvious drawback, however, is the need to manually annotate conversational data; also, the results of the studies are not comparable since they all use different annotation schemas.
We follow up on this line of research and show that it is possible to automatically extract and compare patterns of mixed initiative across different dialogue datasets, including but not limited to the SCSData, MISC, MSDialog and Ubuntu Dialog Corpus.

\subsection{Mixed-Initiative Design}

% mi ds
Mixed-initiative AI systems are specifically designed to enhance human-machine collaboration~\cite{Cohen1998}.
% why recognise switch detect initiative
In particular, a mixed-initiative dialogue system should be able to recognise the user’s cues for initiative switch to provide a response or initiate a discussion when appropriate~\cite{DBLP:journals/umuai/Chu-CarrollB98}.
Therefore, allocation and transfer of initiative, where initiative is defined as control over the direction of the dialogue flow, is at the core of such systems.
Our understanding of mixed initiative in dialogues is important for the design of a conversational search system as a mixed-initiative AI system~\cite{radlinski2017theoretical,thomas2020theories}.

% soa
State-of-the-art research in the context of mixed-initiative design focuses primarily on the selection/generation of clarifying questions~\cite{DBLP:conf/sigir/AliannejadiZCC19,DBLP:conf/www/ZamaniDCBL20}.
\revised{The need for a clarification typically arises in a situation when the user request is ambiguous, and the system should take an initiative to resolve this ambiguity~\cite{DBLP:journals/umuai/SteinGT99}.
Alternatively, the system may also choose to attempt answering the question, even when the user request is ambiguous, and hope to solicit user feedback instead to adjust the answer accordingly.
Asking too many or too few questions is likely to result in ineffective and unnatural dialogues.
Therefore, we resort to a large-scale dialogue analysis to better understand the strategies chosen by human intermediaries in different situations.}

% why detect initiative: when to switch
Measuring mixed initiative is essential for assessing dialogue quality~\cite{Cohen1998}.
However, the standard dialogue evaluation metrics are scoped to accuracy, relevance and grammaticality of a system response~\cite{DBLP:journals/corr/abs-1902-00098,DBLP:journals/corr/abs-1808-07042}.
Our study is designed to fill this gap by introducing a framework for measuring mixed initiative in dialogues.
We demonstrate the utility of this framework by uncovering patterns of initiative across different dialogue types.

% dialogue analysis of mi
The first systematic study of mixed initiative in dialogues is by \citet{DBLP:conf/acl/WalkerW90}.
They perform a manual analysis of dialogue transcripts and study lexical cues, such as the use of anaphora and different utterance types, as a mechanism for switching control in a dialogue.
Their analysis is based on a small sample of dialogues: 24 transcripts from four dialogue datasets across two dialogue types.
They use the approach to utterance type classification and the rules for transfer of control between participants has been proposed by \citet{whittaker1988cues}, which, in turn, is the earliest investigation into initiative and design of mixed-initiative dialogue systems.

\citet{DBLP:conf/acl/WalkerW90} discover different patterns of mixed initiative to be characteristic of different dialogue types.
They observe that dialogue participants interact differently depending on the distribution of knowledge between them.
In the case of advisory dialogues, such as financial consultation and technical support calls, control over the conversation is shared almost equally between an expert-assistant and an information seeker.
In other dialogues, however, an expert-assistant tends to control 90\% of the interactions.
In these dialogues, experts give out instructions unless interrupted by an information seeker requesting additional clarifications.
% what did we do
In this paper, we show how to scale this type of analysis, which has only been performed on a handful of dialogues so far, to thousands of publicly available dialogue transcripts using automated techniques.
Our results cast light on the asymmetry of roles that occurs across different dialogue types, including information-seeking dialogues.

% differences to our earlier work
\revised{In this work, we propose to identify and characterize the general interaction patterns in information-seeking dialogues using the QRFA model~\cite{DBLP:conf/ecir/VakulenkoRCR19}.
QRFA provides a set of coarse utterance labels that allow for a dialogue analysis on a much higher level of abstraction in comparison with the labels used in other frameworks proposed for modeling the structure of information-seeking dialogues, such as COR~\cite{DBLP:journals/ipm/SitterS92} and SCS~\cite{trippas2018informing}.}

Our study builds upon the dialogue analysis experiments reported in our previous work~\cite{DBLP:conf/ecir/VakulenkoRCR19,DBLP:conf/sigir/VakulenkoKR20}.
We extend those experiments in several important directions:
\begin{enumerate*}
\item with an evaluation of the automated utterance classification approach;
\item with a comparison of public dialogue datasets to a sample of virtual reference interviews; and
\item with a comparison of two dialogue analysis approaches introduced in \citep{DBLP:conf/ecir/VakulenkoRCR19,DBLP:conf/sigir/VakulenkoKR20} on a large set of dialogue datasets.
\end{enumerate*}

\section{The ConversationShape Framework}
\label{sec:framework}

We propose a dialogue analysis framework, ConversationShape, that helps to detect patterns of mixed initiative in dialogue transcripts.
As we will show in our dialogue analysis, these patterns are instrumental in identifying and characterising different dialogue types.
ConversationShape includes:
\begin{enumerate}
\item \textit{fingerprinting}, a dialogue representation approach;
\item a \emph{dialogue flow} model that highlights regular patterns of initiative dynamics; and
\item a set of \textit{asymmetry metrics} that reflect the distribution of initiative between dialogue participants.
\end{enumerate}

% how do these three components relate?
\noindent%
Fingerprinting helps to encode basic structural features of a dialogue and to provide an abstraction suitable for a domain-independent dialogue analysis.
We use fingerprints to model dialogue flow and compare dialogue asymmetry across all the datasets listed in Table~\ref{tab:datasets}.
Both dialogue flow diagrams and asymmetry metrics are methods focused on extracting patterns of mixed initiative in dialogue.
Dialogue flow reflects the dynamics of mixed initiative using sequence mining of regular turn switches between the speakers.
Asymmetry metrics reflect the balance of initiative distribution along several dimensions of initiative simultaneously.
We apply both approaches to analyse the content of 16 dialogue datasets, thereby revealing similarities and differences in their dialogue structure (see Section~\ref{sec:results}).

\subsection{Fingerprinting Dialogues}
\label{sec:dialogue_representation}

% More formally
To produce a dialogue representation that can be used for both types of dialogue analysis presented in this paper (dialogue flows and asymmetry measurements), we apply fingerprinting to dialogue transcripts.
We consider a dialogue transcript to consist of a sequence of utterances $D_i = [U_{i1}, U_{i2}, \ldots, U_{ij}, \ldots, U_{in}]$, where each utterance $U_{ij}$ is a text, i.e., a sequence of characters.
Since the original utterances may be long, especially in forum threads (see Table~\ref{tab:datasets}), we split them into sentences to simplify utterance classification.

To produce a dialogue fingerprint, we annotate every utterance $U_{ij}$ with the following features:
\begin{enumerate}
\item speaker \textbf{role}: Seeker or Assistant;
\item utterance \textbf{length}: an integer for the number of characters in $U_{ij}$;
\item utterance \textbf{type}: Hi, Initiative, \revised{NonInitiative}, Bye; and
\item term \textbf{repetitions}: a binary vector that indicates the terms in $U_{ij}$ that are repeated, i.e., encountered more than once, within the same dialogue $D_i$.
\end{enumerate}

% more details approach: overview of the rest of the section
\noindent%
Next, we describe these features and the steps that are necessary to produce a dialogue fingerprint.
At the end of this section, our fingerprinting approach is illustrated using a sample dialogue from the ReDial dataset~\cite{li2018towards}.
%  we also provide details on training and evaluation of our utterance classifier.

% \subsection{Fingerprint structure}

% more details on each of the features
\paragraph{Speaker role}
Information-seeking dialogues often have a clearly defined role for each of the dialogue participants.
One of the participants has an information need (Seeker) and the other one aims to provide information that can satisfy this need (Assistant).

% Some of the grounded datasets are annotated with roles as well: one of the speakers who has access to the background knowledge is considered to be an Assistant. Otherwise, for the chit-chat datasets, where there is no explicit roles assigned to the dialogue participants, we apply the same heuristic as for the Ubuntu dataset.

\paragraph{Utterance length}
An important feature that we want to be reflected in the dialogue fingerprint is the share that each of the participants contributes to the dialogue content.
We calculate the number of characters in each of the utterances.
Below, this will help us to estimate the balance between the dialogue participants in terms of their contributions to the dialogue content.
Other metrics can be used for this purpose as well, such as the number of words, subword tokens, phonemes, or the time taken by each of the dialogue participants in case of spoken dialogue.

\paragraph{Utterance type}

% describe utterance types
To recognise utterances that carry initiative in a dialogue, we assign to every utterance $U_{ij}$ one of the labels $t$ from a closed set $T=\{H, I, N, B\}$.
\revised{Utterance types are assigned independently from the speaker roles.
The labels are assigned as follows.
Types H for `\textit{Hi}' and B for `\textit{Bye}' are used to filter out utterances that express greetings and farewells.
We use type I for \textit{Initiative} to distinguish utterances that include:
\begin{enumerate*}
    \item questions by either of the speaker roles  (\textit{Can you help me find \ldots? Are you looking for \ldots?}); 
    \item statements containing a request (\textit{Please, help me find \ldots Please, provide the following information \ldots}); and
    \item statements describing an information need (\textit{I'm looking for \ldots\ I need \ldots}).
\end{enumerate*}}
All other utterances are considered to be of type N.

To assign a label $t_{ij}$ to every utterance $U_{ij}$, we use a function $F: U_{ij} \mapsto t_{ij}$.
This function $F$ is learned by training a supervised classification model.
For more information on our approach to training and evaluation of the utterance classifier, see Section~\ref{utterance_classifier}.

\paragraph{Term repetitions}

% motivation
To keep track of repetition patterns, we construct a binary matrix that indicates which frequent term appears in which utterance.
A term is considered frequent if it appears in the same dialogue more than once.
We start by converting every utterance into a bag-of-words representation and apply standard pre-processing techniques: remove punctuation, split utterances into words, lowercase, remove stopwords,\footnote{\url{https://raw.githubusercontent.com/stopwords-iso/stopwords-en/master/stopwords-en.txt}} and apply the English Snowball stemmer.
In this manner, every utterance is represented as a set of terms (see Table~\ref{tab:fingerprint} for an example).

Let us call the set of all frequent terms in dialogue $i$, the \emph{dialogue vocabulary} $V_i$.
Term repetitions are stored as a binary matrix, where every row $j$ corresponds to the utterance in the dialogue $U_{ij} \in D_i$ and every column $k$ corresponds to the term in the dialogue vocabulary $w_k \in V_i$.
Every utterance is encoded into a binary vector using the dialogue vocabulary: $v_{ijk} = 1$ if $w_k \in U_{ij}$ else $v_{ijk} = 0$, where $v_{ijk}$ is the element of the dialogue vocabulary matrix for utterance $j$ and term $k$ of $V_i$.
The dialogue vocabulary can be discarded after all term repetitions are stored in the binary matrix.
This approach allows us to easily track repetition patterns in a dialogue: for each term we can identify the speaker who first introduced it and whether it was subsequently reused by another speaker.
\revised{Term repetitions are part of the dialogue fingerprint.}

\theoremstyle{definition}
\begin{definition}[Dialogue fingerprint]
\revised{A dialogue fingerprint $F_i$ is a matrix of size $n_i \times m_i$, where $n_i$ is the number of utterances in dialogue $i$ and $m_i$ is the number of features that represent dialogue $i$.}
We use the following set of features in our dialogue analysis: $r_{ij} \in R$ is the role a participant plays in a dialogue $R=\{S, A\}$, \revised{$t_{ij} \in T$ is one of the utterance types $T=\{H, I, N, B\}$, $l_{ij}$ is the utterance length $l_{ij} = |U_{ij}|$}, and $\textbf{v}_{ij}$ is a vector indicating appearance of the frequent terms in $U_{ij}$.
Thus, each row $j$ of the matrix $F_i$ corresponds to a tuple \revised{$\langle r_{ij}, t_{ij}, l_{ij}, \textbf{v}_{ij}\rangle$} that represents an utterance $U_{ij} \in D_i$.
\end{definition}

\paragraph{Illustrative example}
\label{par:example}

Let us consider a sample dialogue to illustrate how fingerprinting works in practice.
We will use a snippet of a dialogue transcript from the ReDial dataset~\cite{li2018towards} (S stands for Seeker, A for Assistant):

\begin{description}
\item[\textbf{(A)}] Hey! What kind of movies do you like to watch?
\item[\textbf{(S)}] I’m really big on indie romance and dramas
\item[\textbf{(A)}] Ok what's your favorite movie?
\item[\textbf{(A)}] Staying with that genre, have you seen @88487 or @104253
\item[\textbf{(A)}] Those are two really good ones
\item[\textbf{(S)}] When I was a kid I liked horror like @181097
\item[\textbf{(A)}] @Misery is really creepy but really good. I only recently got into horror.
\end{description}

% \noindent%
\begin{table}[t]
\caption{Illustration of the fingerprinting approach for dialogue representation. In this example the dialogue vocabulary consists of two terms that appear more than once in this dialogue: $V_i=\{\textbf{movi}, \textbf{horror}\}$.}
\label{tab:fingerprint}
\resizebox{\columnwidth}{!}{%
\begin{tabular}{ cc r cc l l}
\toprule
\multicolumn{5}{c}{\bf Fingerprint} & \multicolumn{1}{l}{\multirow{2}{*}{\bf Utterance}} & \multicolumn{1}{l}{\multirow{2}{*}{\bf Terms}} \\
\cmidrule{1-5}\textbf{Role} & \textbf{Type} & \textbf{Length} & \multicolumn{2}{c}{\bf Repetitions} & &  \\
\midrule
A & H & 4 & 0 & 0 & Hey! & \{hey\} \\
A & I & 41 & 1 & 0 & What kind of movies do you like to watch? & \{watch, \textbf{movi}\} \\
S & N & 42 & 0 & 0 & I’m really big on indie romance and dramas & \{romanc, drama, indi\} \\
A & I & 30 & 1 & 0 & Ok what's your favorite movie? & \{favorit, \textbf{movi}\} \\
A & I & 56 & 0 & 0 & Staying with that genre, have you seen @88487 or @104253 & \{genr, stay, 88487, 104253\} \\
A & N & 30 & 0 & 0 & Those are two really good ones &  \{\} \\
S & N & 44 & 0 & 1 & When I was a kid I liked horror like @181097 & \{181097, kid, \textbf{horror}\} \\
A & N & 41 & 0 & 0 & @Misery is really creepy but really good.  & \{miseri, creepi\} \\
A & N & 32 & 0 & 1 & I only recently got into horror.  & \{\textbf{horror}\} \\
\bottomrule
\end{tabular}
}
\end{table}

\noindent%
The fingerprint of this dialogue is given in Table~\ref{tab:fingerprint}.
Each row corresponds to an utterance annotated with the features described above.
Note that we further segment the original utterances provided in the dataset using the sentence boundaries to reduce the utterance lengths and make the classification task, which is required for annotating utterance types, easier.
For more details on utterance classification see Section~\ref{utterance_classifier}.

The first column contains the speaker role (A -- Assistant, S -- Seeker).
The second column contains utterance types (H -- Hi; I -- Initiative; N -- NonInitiatiave).
The third column contains utterance lengths (measured as the number of characters).
The last two columns indicate which frequent terms (dialogue vocabulary $V_i = {movi, horror}$) appear in which utterance.
In this way, the dialogue is encoded into a sequence, where every utterance is represented with exactly 5 features: [A-H-4-0-0, A-I-41-1-0, S-N-42-0-0, \ldots, A-N-32-0-1].

This representation is very compact and privacy-preserving.
Since only the structural features of the dialogue are preserved and the dialogue vocabulary is concealed, there is no way to recover the dialogue content from its fingerprint.
However, this representation is sufficient for the two types of dialogue analysis that we present in the next two sections.

\subsection{Dialogue Flow}
\label{sec:dialogue_flow}

% \paragraph{Approach}

% what is a dialogue flow diagram? why?
A dialogue flow diagram allows us to observe how initiative switches between dialogue participants.
\revised{We produce a separate diagram for each of the datasets and use them to compare patterns of mixed initiative in information-seeking dialogues and other types of dialogue.}

% how
To derive a diagram of dialogue flow, we apply sequence mining to dialogue fingerprints.
Since a fingerprint is a matrix we convert it into a single sequence first \revised{using utterance types.
In this way, every dialogue is represented as a sequence of utterance types.}

We focus on the utterance types that indicate dynamics of initiative (Initiative and NonInitiatiave), and extend them to indicate the speaker role as well.
The utterances that correspond to greetings (Hello) and farewells (Bye) are ignored.
The resulting label set corresponds to the QRFA annotation schema previously used by \citet{DBLP:conf/ecir/VakulenkoRCR19}: two labels for the Seeker role (Query and Feedback) and two labels for the Assistant role (Request and Answer).
See Table~\ref{tab:mapping} for the correspondence between our utterance types and the QRFA labels accompanied with sample utterances.

A dialogue flow diagram reflects the counts of all unigrams and bigrams \revised{of utterance types across all dialogue sequences in the dataset}.
The diagrams are constructed using the same template with circles that represent the dialogue start and the dialogue end, and rounded boxes for the QRFA utterance labels.
Boxes represent unigrams and arrows represent bigrams.
The color intensity (opacity) of the arrows and boxes represents proportions of unigrams and bigrams across all QRFA sequences in the dataset.

\begin{table}[t]
\caption{Mapping from a dialogue fingerprint to QRFA schema reflecting patterns of initiative switch between dialogue participants. Sample utterances were selected from different dialogue datasets using automatic annotations produced by our utterance type classifier.}
\label{tab:mapping}
\resizebox{\columnwidth}{!}{
\begin{tabular}{llll}
\toprule
\bf Role & \bf Type & \bf QRFA & \bf Sample Utterances \\
\midrule
Seeker & Initiative & Question &
\begin{tabular}[c]{@{}l@{}}What was the relationship between Coolio and Gangsta's paradise?\\ I am looking for information about the City Centre North B and B hotel.\\ Trying to find a pdf of: Adv Exp Med Biol.\end{tabular} \\
\midrule
Assistant & Initiative & Request & \begin{tabular}[c]{@{}l@{}}Which Graphic card is installed on your machine?\\ Would you like to try a different departure station?\\ Is that what you are looking for?\end{tabular} \\
\midrule
Seeker & NonInitiatiave & Feedback & \begin{tabular}[c]{@{}l@{}}Yes\\ I like the acting the actors that were in it.\\ Thanks a lot for your recommendations!\end{tabular} \\
\midrule
Assistant & NonInitiatiave & Answer & \begin{tabular}[c]{@{}l@{}}I will try to look it up\\ It is a language program that teaches speaking and understanding the language.\\ There are 5 guesthouses that have free parking.\end{tabular}\\
\bottomrule
\end{tabular}
}
\end{table}

The opacity of a box in the diagram corresponds to the respective unigram count normalised by the total count of all unigrams, e.g., the opacity of the $Q$ box is $\frac{\#Q}{\#Q + \#A + \#R + \#F} \times 100\%$.
To calculate the opacity for arrows that connect boxes to circles, i.e., the beginning and end of the dialogue, we use bigram counts by prepending < and appending > \revised{characters} to every sequence, and then normalise them by the number of dialogues, \revised{e.g., the opacity of the arrow from the start node to $Q$ is $\frac{\#<Q}{d} \times 100\%$, where $d$ is the number of dialogues in the dataset}.
The opacity of arrows reflects the proportion of times a sequence begins or ends with the corresponding label.

For example, in the QuAC dataset illustrated in Figure~\ref{fig:qa} (see Section~\ref{subsection:evaluationofutteranceclassification}), half of the utterances are questions (the opacity of the box labeled $Q$ is 50\%) and half of the utterances are answers (the opacity of the box labeled $A$ is 50\%).
This is unsurprising since QuAC is a conversational QA dataset and the data collection task setup has fixed the dialogue structure in advance.
% All questions are followed by answers (the opacity of the QA arrow is 100\%) but not all answers are followed by questions (the opacity of the AQ arrow is 86\%) since every dialogue comes to an end.
The only possible start of the dialogue is a question by the Seeker (the opacity of the <Q arrow is 100\%), and the only possible end of the dialogue is an answer by the Assistant (the opacity of the A> arrow is 100\%).
\revised{The opacities for the unigrams always sum to 100\% as do the opacities of the starting arrows and the ending arrows. However, this approach does not work for all other bigram counts.}

% Therefore,
\revised{The same bigram may repeat multiple times in the same dialogue.
In this case, the total count will be higher than the number of dialogues.
For example, if every dialogue contains two question-answer pairs then the opacity of the QA arrow should be 200\%.
Alternatively, if we calculate the arrow opacity as the relative proportion of all bigrams (same as for unigrams), most of the arrows will be hardly visible and to tell the difference between their relative proportion will be rather difficult.
For example, in a balanced dialogue the fraction of occurrences of each of the eight bigrams QA, AQ, FA, AF, QR, RQ, FR, and RF is equal, $1/8=12.5\%$.}

\revised{To normalise the opacity, we divide every bigram count by the maximum bigram count.
The maximum bigram count is obtained by counting all bigrams and then taking the maximum number in this set. For example, if the most common bigram was QA with count 230 the opacity of the QA arrow will be 100\% while the opacity of the AQ arrow, assuming its count was 23, will be 10\%.}
% rather than by the total count

% \revised{Since dialogues often span across multiple turns, . In case of unigrams (and the bigrams starting with < or ending with >) we have only four classes in total (QRFA), in case of bigrams we have 8 classes (12 permutations in total minus 4 that occur within the turns, i.e., between QF and RA). The relative \% for bigrams is much lower than for unigrams and the difference between the color opacity is less apparent, i.e., won't be visible on the dialogue flow diagram. That's why we use relative frequency with respect to the maximum bigram count to compare the proportion of different bigrams.}
% subsequences often repeat and the number of bigrams is very high.
% This approach reflects the distribution of bigram counts across sequences.
% \revised{We can not use the same approach here as for unigrams since}
% \todo{example for bigrams, to illustrate the effect of normalization by maximum bigram count, rather than total bigram count}

To discard the effect of long turns on the bigram counts (this is especially prominent for  forum threads and for transcripts of spoken dialogues), we reduce sequences to a single label per turn.
This label indicates whether any of the annotated utterances carries initiative (Q and R) or not (F and A).
For example, the fingerprint of the sample dialogue illustrated in Table~\ref{tab:fingerprint} will result in the following sequence of initiative switches: RFRFA \revised{instead of RFR\textbf{A}FA}.
In this way we focus on the initiative switch between dialogue participants and ignore the patterns of initiative within the turn of a single participant.
Therefore, there are no arrows between the utterance labels of the same role in the dialogue flow diagrams, i.e., between Q and F, or R and A.

\subsection{Asymmetry Metrics}
\label{sec:asymmetry_measures}
As a complementary analysis to dialogue flow analysis, we propose a set of metrics that reflects other dimensions of initiative beyond utterance types.
This set of metrics is designed to measure the distribution of initiative between dialogue participants.
% rather than frequent patterns of initiative switch as in dialogue flow diagrams. 
We use \revised{four} metrics to measure initiative in dialogue:
\begin{enumerate}
	\item Volume -- who talks more in a dialogue?
	\item Direction -- who requests information in a dialogue? 
	\item Information -- who contributes to the dialogue topic? and 
	\item Repetition -- who follows up on the topic introduced by another participant?
% 	\item Flow -- who introduces more topics than follows up on the topics introduced by another participant? 
\end{enumerate}
These metrics are derived from the information stored in a dialogue fingerprint. Volume is based on the combination of Role and Length features, Direction is based on the combination of Role and Type, both Information and Repetition metrics are based on the combination of Role and Repetitions. To compare the metrics between dialogues of different lengths we normalise the scores by the number of utterances, i.e., the number of rows in the dialogue fingerprint \revised{$n_i$}.

\emph{Volume} is associated with explicitly seizing control over the dialogue, effectively turning it into a monologue in extreme cases. We estimate Volume by counting the average number of characters for each of the dialogue participants separately. This is achieved by grouping the values of utterance Length from the dialogue fingerprint by Role and summing them up. For every dialogue $D_i$ the Volume for the role $r \in R$ is computed as:
\begin{equation}
  \mathit{Volume_{ir}} = \frac{1}{n_i}\sum _{j=1}^{n_i} l_{ij} [r_{ij}=r].
\end{equation}

\emph{Direction} represents an explicit attempt at controlling the direction of the topic of a dialogue. A question, for example, sets an expectation for another participant to produce a relevant answer. This is achieved by counting the number of utterances with Type \textit{Initiative} grouped by Role. This feature set was used previously for generating dialogue flow diagrams as well. For every dialogue $D_i$ the Direction for the role $r \in R$ is computed as:
\begin{equation}
	\mathit{Direction_{ir}} = \frac{1}{n_i}\sum _{j=1}^{n_i} \mathcal{I}(t_{ij}=I) [r_{ij}=r].
\end{equation}

%  As a reminder, Repetitions is a binary matrix, which is part of the dialogue fingerprint. This matrix indicates which terms appear in which utterance (utterances correspond to the rows of the matrix), similar to the bag-of-words representation. We retain only those terms that are repeated in the dialogue, i.e., occur more than once. Hence, the sum for each column in the Repetitions matrix is > 1.

\emph{Information} reflects the contribution that a participant makes to the dialogue topic. It is derived from the analysis of the repetition patterns (a binary matrix in the dialogue fingerprint).
\revised{This metric is motivated by the term frequency counts, which are often used to determine the dialogue topic.
We estimate Information by counting the number of frequent terms that were coined by each of the dialogue participants.} This is achieved by counting the number of columns in the Repetitions matrix with the first non-zero element pointing to the participant who introduced the term in the dialogue first:
\begin{equation}
   \mathit{Information_{ir}} = \frac{1}{n_i}\sum _{j=1}^{n_i}\sum _{k=1}^{m_i} \mathcal{I}(v_{ijk}=1) [r_{ij}=r] \times \mathcal{I}\left(\sum _{g=1}^{j-1}v_{igk}=0\right).
\end{equation}

\revised{\emph{Repetition} indicates an explicit follow-up on the topic introduced by another participant. It is measured by counting the non-zero elements in the Repetitions matrix for terms that were not introduced by the speaker. We consider repetition as a type of relevance feedback in the dialogue. Term repetition effectively contributes towards the increase of the term frequency counts. Hence, by repeating the term (following up) a speaker implicitly endorses the contribution of the other dialogue participant and increases its importance with respect to the dialogue topic:}
\revised{\begin{equation}
  \mathit{Repetition_{ir}} = \frac{1}{n_i}\sum _{j=1}^{n_i}\sum _{k=1}^{m_i} \mathcal{I}(v_{ijk}=1)[r_{ij}=r] \times \mathcal{I}\left(\sum _{g=1}^{j-1}v_{igk} [r_{ij} \neq r] >0\right).
\end{equation}}

% \emph{Flow} is the difference between Repetition and Information, which reflects on the role of a participant in maintaining the coherence of the dialogue by either referencing previous statements or driving the dialogue forward by introducing new information. Flow is calculated based on the other two metrics, e.g., for the role $r \in R$:
% \begin{equation}
%     \Delta \mathit{Flow_{ir}} = \mathit{Repetition}_{ir} - \mathit{Information}_{ir}.
% \end{equation}

\noindent%
Thus, for every dialogue $D_i$ we measure Volume, Direction, Information and Repetition separately for each of the roles: $\mathit{Metric}_{iA}$ and $\mathit{Metric}_{iS}$, where \emph{Metric} denotes one of the metrics that we have just introduced. Next, we use the average and the difference between $\mathit{Metric}_{iA}$ and $\mathit{Metric}_{iS}$ to compute the averages between all dialogues in the dataset and compare these metrics across different datasets.

% The average between $\mathit{Variable}_{iA}$ and $\mathit{Variable}_{iS}$ shows the magnitude of initiative for each of the metrics, i.e., how much Volume, Direction and Information is expressed by both of the participants in different datasets:
% \begin{equation}
% \mathit{Variable} = \frac{1}{d}\sum _{i=1}^{d} \frac{\mathit{Variable}_{iA} + \mathit{Variable}_{iS}}{2}.
% \end{equation}

The difference between $\mathit{Metric}_{iA}$ and $\mathit{Metric}_{iS}$ allows us to compare the distribution (that is, the balance of initiative) between the dialogue participants.
% A simple way to compare the difference between variables is a fraction that can be defined for each of the dialogue participants separately (S and A), e.g.:
% \begin{equation}
%   \mathit{\%Variable_A} = \frac{1}{d}\sum _{i=1}^{d} \frac{\mathit{Variable}_{iA}}{\mathit{Variable}_{iA} + \mathit{Variable}_{iS}}.
% \end{equation}
% While $\mathit{\%Variable}$ is easy to interpret, we need to keep in mind that 0.5 in this case indicates the exact balance of initiative between the dialogue participants.
To produce a score in the range $[-1; 1]$, with $0$ indicating the exact balance of initiative, $1$ and $-1$ indicating dominance of initiative by either Assistant or Seeker, respectively, we calculate $\Delta \mathit{Metric}$ \revised{across all $d$ dialogues in the dataset} as follows:
\begin{equation}
\Delta \mathit{Metric} = \frac{1}{d}\sum _{i=1}^{d} \frac{\mathit{Metric}_{iA} - \mathit{Metric}_{iS}}{\mathit{Metric}_{iA} + \mathit{Metric}_{iS}}.
\end{equation}

% Thus, we show that by quantifying different aspects of mixed initiative the asymmetry metrics we introduced allow to better interpret the speaker roles and draw conclusion about the nature of the interaction. In the next sections we apply both types of analysis, dialogue flow and asymmetry metrics, to 16 dialogue datasets to uncover relations between different dialogue types based on the patterns of mixed-initiative interaction they exhibit.

\paragraph{Illustrative example.}
We show how the asymmetry metrics are computed using the sample dialogue fingerprint from Table~\ref{tab:fingerprint}.
% Note that we consider all utterances and all columns of the fingerprint here, not only the I and N types as in the dialogue flow analysis described in the previous subsection.
The Assistant clearly dominates the dialogue by the number of utterances and their relative length. \revised{This also leads to a difference between the number of characters ascribed to each of the dialogue participants, which is reflected in the $\Delta \mathit{Volume}$ metric}:
\begin{align}
\mathit{Volume_{iA}} & = \frac{4+41+30+56+30+41+32}{9} \approx 42 \\
\mathit{Volume_{iS}} & = \frac{42+44}{9} \approx 10 \\
% \mathit{Volume} & = \frac{42 + 10}{2} = 26 \\
\Delta \mathit{Volume} & = \frac{42 - 10}{42 + 10} \approx 0.62.
\end{align}
Moreover, all the questions in this dialogue originate from the Assistant. This information is reflected in $\Delta \mathit{Direction}$ metric:
\begin{align}
\mathit{Direction_{iA}} & = \frac{3}{9} \approx 0.33 \\
\mathit{Direction_{iS}} & = 0 \\
% \mathit{Direction} & = \frac{0.33 + 0}{2} = 0.165 \\
\Delta \mathit{Direction} & = \frac{0.33 - 0}{0.33 + 0} = 1.
\end{align}
Both dialogue participants contribute to the dialogue topic. The Assistant was the first to introduce \textit{movies} as the dialogue topic. Later, the Seeker introduced \textit{horror} as a subtopic:
% and the Assistant followed up on it further. In this manner, the Assistant demonstrates the ability to lead the dialogue as well as a readiness to follow up on a topic of interest introduced by the Seeker:
%
\begin{align}
\mathit{Information_{iA}} & = \mathit{Information_{iS}} = \frac{1}{9} \approx 0.11 \\
% \mathit{Information} & = \frac{0.11 + 0.11}{2} = 0.11 \\
\Delta \mathit{Information} & = \frac{0.11 - 0.11}{0.11 + 0.11} = 0.
% \mathit{Repetition_{iA}} & = \frac{1}{9} \approx 0.11 \\
% \mathit{Repetition_{iS}} & = 0 \\
% \mathit{Repetition} & = \frac{0.11 + 0}{2} = 0.055 \\
% \Delta \mathit{Repetition} & = \frac{0.11 - 0}{0.11 + 0} = 1 
\end{align}

\revised{The Assistant followed up on the topic introduced by the Seeker, also repeating the word \textit{horror} that was introduced by the Seeker.
Thereby, the Assistant demonstrates the ability to lead the dialogue as well as the readiness to follow up on the topic of interest introduced by the Seeker:
\begin{align}
% \mathit{Information_{iA}} & = \mathit{Information_{iS}} = \frac{1}{9} \approx 0.11 \\
% % \mathit{Information} & = \frac{0.11 + 0.11}{2} = 0.11 \\
% \Delta \mathit{Information} & = \frac{0.11 - 0.11}{0.11 + 0.11} = 0 \\
\mathit{Repetition_{iA}} & = \frac{1}{9} \approx 0.11 \\
\mathit{Repetition_{iS}} & = 0 \\
% \mathit{Repetition} & = \frac{0.11 + 0}{2} = 0.055 \\
\Delta \mathit{Repetition} & = \frac{0.11 - 0}{0.11 + 0} = 1.
\end{align}
}

\section{Experimental Setup}
\label{sec:experimental_setup}

In this section we describe the setup used for our large-scale dialogue analysis.
We list all the datasets considered in our analysis and point out the relations between them that are known a priori.
In this section we also provide details on training and evaluation of the utterance classification model, which we use to annotate utterance types in all datasets as part of the dialogue fingerprints.

\subsection{Dialogue Datasets}
\label{sec:datasets}

% overview datasets
Our analysis is focused on the information-seeking dialogue datasets that have previously been introduced or analysed in the context of conversational search.
We complement them with datasets for other dialogue types indicated in Figure~\ref{fig:survey}: conversational QA, task-oriented, grounded and chit-chat dialogues.

% introducing the table
Table~\ref{tab:datasets} groups the datasets by type and summarises their main characteristics.
We differentiate dialogues by domain, modality (text or speech), source (forum, chat, task or book), and setup (natural or simulated information need).
Table~\ref{tab:datasets} also provides basic statistics for each of the datasets in terms of the number of dialogues, the dialogue and utterance lengths, and the average number of utterances per turn.

% describe basic stats
Most of the datasets have been collected in an artificial setting (see the value \textit{simulated} under Setup in Table~\ref{tab:datasets}), i.e., the participants were instructed to interact with each other or with a chatbot (Meena/Mitsuku and Meena/Meena).
These interactions are usually mediated by a text-based chat interface (see the value \textit{text} for Modality and \textit{chat} for Source in Table~\ref{tab:datasets}).

\subsubsection{Information-seeking dialogues}

% introduce subtypes of information-seeking dialogues: natural
There is a subset of real information-seeking dialogues that occurred on-line without intrusion of researchers (see Setup \textit{natural} in Table~\ref{tab:datasets}).
These datasets have either been extracted from Q\&A forums or from on-line chatrooms.
We refer to this subset of information-seeking dialogue datasets collected in natural settings as IN, and the rest of the information-seeking dialogue datasets collected in the simulated setting as IS.

% forums
\textit{Q\&A forums} (IN).
MSDialog~\cite{InforSeek_Response_Ranking} has been collected on a technical support forum provided by Microsoft, and MANTIS~\cite{penha2019introducing} is from the community question-answering portal Stack Exchange.
As is evident from Table~\ref{tab:datasets}, both datasets have very different basic statistics from the other dialogue datasets: forum threads are usually much shorter than text- and speech-based dialogues and their utterances are much longer.

% text chat virtual reference interview
\textit{On-line chats} (IN).
Unlike forums, on-line chatrooms provide an opportunity for real-time synchronous communication, which enables a more dynamic interaction that more closely resembles a human dialogue.
In particular, a reference interview with a librarian, which is a classic example of an information-seeking dialogue, has been replaced with a virtual reference interview that occurs in a private chatroom~\cite{radford2013not}.

We obtained a sample of dialogue transcripts of virtual reference interviews from OCLC, which is a non-profit global library cooperative.
This is a random sample with 560 anonymised transcripts of virtual reference interviews, which took place from June 2010 through December 2010.
The OCLC dataset contains real information-seeking dialogues mediated by professional librarians, who were trained to provide reference services.

The most similar dataset of this type that is publicly available, is the Ubuntu Dialogue Corpus~\cite{DBLP:conf/sigdial/LowePSP15}.
It contains dialogue transcripts extracted from a public chat of the Ubuntu user community.
The goal of this chat is to provide technical support and advice on software-related issues.
Note that the dialogues in this dataset are more informal than those in the OCLC dataset since they occur between peers rather than between a customer and a professional service provider.

% spoken
\textit{Spoken Conversational Search} (IS).
The SCSdata~\cite{trippas2018informing} and MISC~\cite{thomas2017misc} datasets resulted from two separate laboratory data collection tasks, in which pairs of human volunteers interacted in the context of an information-seeking task.
One of the volunteers had access to the textual description of the information need, and the other one was using a search engine to find relevant information that can satisfy this need. 
Their spoken dialogues were recorded and then transcribed, either manually in the case of SCSdata or automatically in the case of MISC.
 
% conversational recommendation
\textit{Conversational recommendation} (IS) shares a number of similarities with the conversational search task and the boundaries between the two are not explicitly defined~\cite{jannach2020survey,gao-2021-advances-arxiv}.
Thus, we consider dialogues collected to inform the conversational recommendation task to constitute information-seeking dialogues.
Both datasets we consider in the context of conversational recommendation, ReDial~\cite{li2018towards} and CCPE~\cite{radlinski2019coached}, focus on the movie recommendation scenario.
While ReDial contains dialogues in which human participants were instructed to recommend movies to watch, CCPE is focused specifically on the preference elicitation phase, in which the assistant is instructed to learn more about the user tastes rather than make a recommendation.

% Qulac and QuAC
\textit{Conversational QA} (IS) datasets result from task designs that fix the structure of an information-seeking conversation in advance, following a pre-defined template.
Therefore, such datasets are not suitable for discovering and analysing the structure of naturally occurring dialogues.
For example, the QuAC dataset~\cite{DBLP:conf/emnlp/ChoiHIYYCLZ18} consists of dialogues that contain sequences of question-answer pairs.
The Qulac dataset~\cite{DBLP:conf/sigir/AliannejadiZCC19} contains synthetic dialogues of length 3, in which every user question is followed by a clarifying question and a user response.
Since the utterance types are known a priori, we use this dataset to train our utterance classification model.

% introduce other dialogue types
\subsubsection{Other dialogue types}

The set of information-seeking dialogues is complemented with other dialogue datasets collected to train and evaluate models for conversational QA, task-oriented, grounded and chit-chat dialogues.

% task-oriented
\textit{Task-oriented} (TO) dialogues aim to accomplish a certain task or a sequence of related subtasks, such as booking a restaurant and calling a cab.
We include the MultiWOZ dataset~\cite{DBLP:conf/emnlp/BudzianowskiWTC18} in our analysis to discover the typical structure of a task-oriented dialogue and compare it with the structure of an information-seeking dialogue.
MultiWOZ is a large-scale dataset that spans across multiple domains and topics.
While we include a single task-oriented dataset in our analysis, we consider it representative of this dialogue type due to its size (10K dialogues is an order of magnitude larger than all previous annotated task-oriented corpora~\cite{DBLP:conf/emnlp/BudzianowskiWTC18}) and diversity (single-domain and multi-domain dialogues about restaurants, hotels, attractions, taxi, trains, hospitals and police).

% chit-chat
\textit{Chit-chat} (CC) dialogues are used to design and evaluate social chatbots~\cite{INR-074}.
The primary goal of a social chatbot is to entertain the user by holding a human-like conversation.
The Meena dataset consists of three subsets (Meena/Mitsuku, Meena/Meena and Meena/Human) used for chatbot evaluation~\cite{DBLP:journals/corr/abs-2001-09977}.
Two of them (Meena/Mitsuku and Meena/Meena) contain transcripts of human-machine dialogues produced by volunteers interacting with two social chatbots.
The third subset (Meena/Human) contains transcripts of human-human dialogues that follow the same setup and are used as a reference point for chatbot evaluation.
The DailyDialog dataset~\cite{DBLP:journals/corr/abs-1710-03957} contains samples of dialogues that occur in common daily situations, such as shopping, doctor appointment etc.
It is different from all other datasets listed in Table~\ref{tab:datasets} since these dialogues were extracted from textbooks for English learners.
This dataset is commonly considered as chit-chat~\cite{sinha2020learning}.

% grounded
\textit{Knowledge-grounded} (KG) dialogues are similar to chit-chat but the additional goal is also to communicate certain information (knowledge).
This knowledge is provided as input (grounding) to the dialogue model either as text (PersonaH~\cite{see2019what} and WoW~\cite{dinan2018wizard}) or a knowledge graph (OpenDialKG~\cite{DBLP:conf/acl/MoonSKS19}).
For example, the PersonaH dataset contains human-human dialogues grounded in short text snippets.
These snippets contain descriptions of personal information, such as hobby and occupation, that the participants are supposed to use in their replies~\cite{see2019what}.
Every dialogue in the WoW dataset revolves around a topic discussed in one of the Wikipedia articles and dialogues in the OpenDialKG dataset use facts stored in the Freebase knowledge graph.

% \todo{@SV check dataset descriptions: 1+2 Q\&A/support forums; 3+4 on-line chat; 5+6 SCS=Spoken Conversational Search; 7+8 conv rec; 9+10 conv qa; 11 multi-domain; chit-chat; grounded }

\begin{sidewaystable}
\centering
\mbox{}\vspace*{11.5cm}
\caption{16 dialogue datasets used in our analysis: 15 public and 1 private (OCLC). The Meena dataset consists of three subsets, which contain human-human and human-machine dialogues (*) that were collected for a chatbot evaluation. All other dialogue datasets contain only human-human dialogues. Dial. len.\ -- average dialogue length calculated as the number of utterances. Utt. len.\ -- average utterance length calculated as the number of words. Utt./turn -- average number of utterances per single dialogue turn produced by one of the dialogue participants. The number in brackets is a standard deviation from the mean. Outliers for the different dimensions are highlighted in \textbf{bold}.}
\label{tab:datasets}
\setlength{\tabcolsep}{4pt}
\begin{tabular}{l r ll c lll rrr}
    \toprule
\bf Dataset & \bf Dialogues & \bf Domain & \bf Type & \bf Subtype & \bf Modality & \bf Source & \bf Setup & \bf Dial. len. & \bf Utt. len. & \bf Utt./turn \\
    \midrule
MSDialog~\cite{InforSeek_Response_Ranking} & 35,500 & tech & info-seek & IN & text & forum & \textbf{natural} & 9 (25) & \textbf{67 (87)} & 1 (3) \\
MANTIS~\cite{penha2019introducing} & 1,400 & multi & info-seek & IN & text & forum & \textbf{natural} & 4 (1) & \textbf{98 (160)} & 1 (0) \\
OCLC\footnote{The dataset is an intellectual property of OCLC \url{https://www.oclc.org} and is subject to a licence agreement.} & 560 & library & info-seek & IN & text & chat & \textbf{natural} & 25 (20) & 12 (15) & 1 (1) \\
Ubuntu~\cite{DBLP:conf/sigdial/LowePSP15} & 1,200,000 & tech & info-seek & IN & text & chat & \textbf{natural} & 6 (8) & 10 (9) & 1 (1) \\
SCSdata~\cite{trippas2018informing} & 37 & web & info-seek & IS & \textbf{speech} & chat & simulated & 27 (21) & 16 (25) & 1 (0) \\
MISC~\cite{thomas2017misc} & 110 & web & info-seek & IS & \textbf{speech} & chat & simulated & \textbf{120 (47)} & 7 (7) & \textbf{2 (2)} \\
ReDial~\cite{li2018towards} & 10,000 & movies & info-seek & IS & text & chat & simulated & 18 (5) & 6 (5) & 1 (0) \\
CCPE~\cite{radlinski2019coached} & 502 & movies & info-seek & IS & text & chat & simulated & 23 (6) & 12 (13) & 1 (0) \\
Qulac~\cite{DBLP:conf/sigir/AliannejadiZCC19} & 10,277 & web & info-seek & IS & text & \textbf{task}\footnote{In Qulac, participants were not paired to participate in a live conversation but added their responses into an on-line form given the previous utterance and the information need description as a prompt.} & simulated & 3 (0) & 8 (3) & 1 (0) \\
QuAC~\cite{DBLP:conf/emnlp/ChoiHIYYCLZ18} & 11,600 & Wikipedia & info-seek & IS & text & chat & simulated & 14 (4) & 9 (7) & 1 (0) \\
MultiWOZ~\cite{DBLP:conf/emnlp/BudzianowskiWTC18} & 10,000 & multi & \textbf{task-orient} & TO & text & chat & simulated & 13 (5) & 13 (6) & 1 (0) \\
Meena/Mitsuku*~\cite{DBLP:journals/corr/abs-2001-09977} & 100 & open & social & CC & text & chat & simulated & 18 (4) & 8 (10) & 1 (0) \\
Meena/Meena*~\cite{DBLP:journals/corr/abs-2001-09977} & 91 & open & social & CC & text & chat & simulated & 19 (5) & 6 (4) & 1 (0) \\
Meena/Human~\cite{DBLP:journals/corr/abs-2001-09977} & 95 & open & social & CC & text & chat & simulated & 15 (2) & 13 (10) & 1 (0) \\
DailyDialog~\cite{DBLP:journals/corr/abs-1710-03957} & 11,000 & multi & social & CC & text & \textbf{book}\footnote{These dialogues were extracted from a textbook for English learners, i.e., likely created by a single author.} & simulated & 7 (4) & 13 (10) & 1 (0) \\
PersonaH~\cite{see2019what} & 102 & personal & social & KG & text & chat & simulated & 12 (0) & 9 (4) & 1 (0) \\
WoW~\cite{dinan2018wizard} & 22,000 & Wikipedia & social & KG & text & chat & simulated & 9 (1) & 16 (7) & 1 (0) \\
OpenDialKG~\cite{DBLP:conf/acl/MoonSKS19} & 13,800 & Freebase & social & KG & text & chat & simulated & 6 (2) & 12 (6) & 1 (0) \\
  \bottomrule
\end{tabular}
\end{sidewaystable}

\revised{All information-seeking datasets have turns annotated with the participant roles, except for the Ubuntu dataset.
Since the dialogues were automatically extracted from a public chatroom, the turns in Ubuntu are annotated only with the user handles.
We use a simple heuristic and assume the first speaker, who initiates the dialogue, to be the Seeker. This decision is motivated by the observation that the Seeker is more likely to initiate the conversation to indicate the information need.
All dialogues in OCLC and MSDialog are also initiated by the Seeker.
We apply the same heuristic to assign the roles in social dialogues to show that it provides for a random assignment if the participant roles are symmetric.}

\subsection{Utterance Classification Model}
\label{utterance_classifier}

% training details
We train our utterance classification model by fine-tuning a pre-trained RoBERTa base with 12 hidden layers and 12 attention heads~\cite{DBLP:journals/corr/abs-1907-11692}.
We use a combination of four datasets to train our model: QuAC, SPAADIA~\cite{leech2013spaadia}, Qulac and NPS chat~\cite{DBLP:conf/semco/ForsythandM07}.

QuAC contains question-answer pairs, which we use as samples for Inititative (I) and NonInitiative (N) utterance types, respectively. 
From the NPS chat corpus, we obtain additional examples of questions (I), greetings (H) and farewells (B).

Qulac and SPAADIA are used to train the classifier to recognise those information requests that are not formulated as questions.
\revised{Every sample in Qulac contains a topic description (e.g., ``Find a dieting advice''), a facet description (e.g., ``Find crash diet plans''), a clarifying question (e.g., ``do you want to know if dieting is safe''), and an answer (e.g., ``no I don't.'')}
We use questions, topic and facet descriptions to train I, and answers to train N type.

The SPAADIA dataset provides examples of the four main sentence types: declarative (statements), imperative (commands), exclamatory (exclamations) and interrogative (questions).
We use declarative sentences to train the N type; imperative and interrogative sentences for the I type.

The resulting dataset contains 86K utterance-labels pairs (H: 1.4K, I: 42K, N: 42K, B: 195).
We randomly split it into two subsets: 77.5K for training and 8.6K for testing our classifier.

% This model achieves micro-average F1 of 0.996 on the held-out test set.

% % preprocessing
% Note the difference in average utterance lengths in Table~\ref{tab:datasets}, which is especially pronounced for the Q\&A forum datasets.
% Messages in forums often contain long texts spanning across several sentences of different types, including declarative, imperative and interrogative.
% Since our classification model was trained on the chat corpora with relatively short utterances, we split all utterances into sentences using punctuation before classification.
% This approach allows us to normalise average utterance lengths across different dataset types.
\subsection{Evaluation of Utterance Classification}
\label{subsection:evaluationofutteranceclassification}
% 0.942457 0.996
Our utterance classification approach achieves a macro-average F1 score of 0.942 on the held-out test set.
Then, we use this classification model to annotate utterances in all the datasets listed in Table~\ref{tab:datasets}.
The Ubuntu dataset is too large even for an automated utterance classification (more than 1M dialogues).
To save computational resources, we annotated a subset of \revised{the first} 50K dialogues and consider it to serve as a representative sample for our dialogue analysis.

\begin{figure}[t]
  \centering
  \includegraphics[width=0.45\linewidth,fbox]{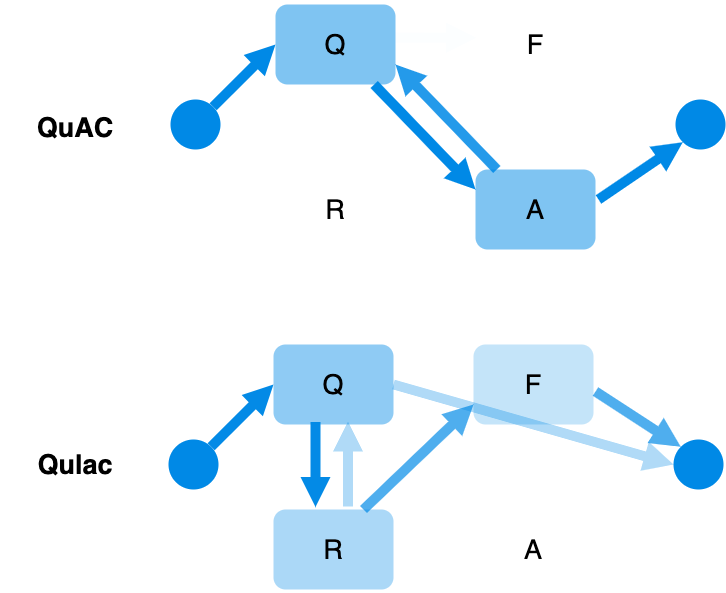}
  \includegraphics[width=0.10\linewidth]{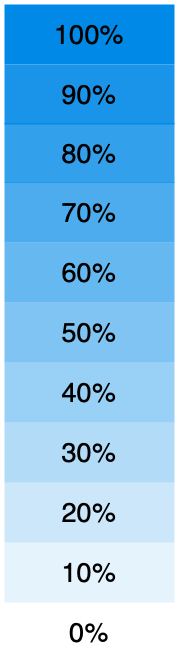}
  \caption{Models of dialogue flow extracted from QuAC and Qulac datasets using automatically annotated utterance labels. \revised{The vertical bar indicates the correspondence between the opacity levels for boxes/arrows and the unigram/bigram frequencies, respectively. This color coding schema is consistent across all the dialogue flow models in the following sections.}}
  \label{fig:qa}
\end{figure} 

\revised{We performed an error analysis by manually inspecting the results.
Two of the paper authors labeled 1,494 randomly selected utterances independently.
We made sure to keep the balance of classes in the random sample we selected (half was labeled as Initiative by our classifier and half as NonInitiative).
There were only 12 cases of disagreement, in total (99\% utterances received the same label by both annotators).}
% a sample of utterances labeled as Initiative to verify that the model learned to recognise different types of utterances.

\revised{The accuracy of our utterance classifier on this random subset is 92\% (113 utterances were classified incorrectly).
Here are some examples of utterances that were labeled as Initiative by our utterance classifier, which demonstrate that the classification model also learned to recognise different formulations for information requests beyond questions: (1) ``\textit{I have found the files, but what am I supposed to use to open them?}’’ (2) ``\textit{\textit{Tell me about} your favorite movie.}’’ (3) ``\textit{\textit{I'm looking for} some suggestions for good movies.}’’}

% how do we evaluate labelling approach
To verify the impact of the utterance classification on the results of the dialogue flow analysis, we produced dialogue flow diagrams for the QuAC and Qulac datasets using automatic annotations (see Figure~\ref{fig:qa}).
While the dialogue flow diagram for the QuAC dataset perfectly fits our assumption about these dialogues as a sequence of question-answer pairs, the triples in the Qulac dataset turned out to be more diverse than we initially thought.
In some cases (31\% of the dialogues) the Seeker replied to a clarifying question with feedback paired with a follow-up question.
Examples: ``yes specifically how is it different from hdl and vldl'', ``yes and also how would i apply'', ``can you just show me the human society's homepage.''
By manually examining such cases, we conclude that our utterance classifier successfully learned to distinguish questions erroneously annotated as \revised{NonInitiative statements} in the training set, i.e., the classifier learned to assign correct labels with overfitting the training data.

To extend the evaluation of our utterance classifier to other dialogue datasets that were not used during training, we utilise the manual labels provided along with the SCSdata and MSDialog-Intent datasets.
MSDialog-Intent is a subset of MSDialog with 2,199 dialogues with utterances manually annotated by crowd workers.

SCSdata and MSDialog-Intent were annotated with two disjoint label sets.
Utterances in SCSdata were manually annotated with 83 labels, called actions, such as ``Info about document’’, ``Query repeat’’, ``Performance feedback’’, etc.
The MSDialog annotators used a set of 12 labels, such as ``Original Question’’, ``Clarifying Question’’, ``Potential Answer’’ and ``Positive Feedback’’.
Therefore, we had to manually map both label sets to the QRFA labels.

Table~\ref{tab:scs_msdialog} shows the mapping that we established between the manually annotated labels and the QRFA schema.
The table also contains examples of utterances extracted from both of the datasets for each of the QRFA labels.
% Thereby, all utterances with labels for commands, questions and requests are considered as Initiative (Q or R), and all other utterance labels, such as answers and feedback subtypes, are considered as NonInitiatiave (F or A).

Our utterance labelling approach achieved a micro-average F1 score of 0.8 on the SCSdata dataset and only 0.17 on the MSDialog-Intent dataset.
84\% of the errors for the SCSdata dataset are due to the inability of the model to recognise Initiative.
In contrast, 70\% of the errors on the MSDialog-Intent dataset are due to the model predicting Initiative where it was not annotated by the crowd workers.

By manually examining all errors in the SCSdata dataset and a random sample of errors from the MSDialog-Intent dataset, we identify two main reasons behind these misclassification results: (1)~utterances annotated as Initiative do not contain an explicit question (see ``Original Question'', ``Intent clarification'' and ``Asks to repeat'' in Table~\ref{tab:scs_msdialog}); and (2)~utterances that contain explicit questions or requests were not annotated as such (see ``Further Details'', ``Potential Answer'',  ``SERP with modification + Interpretation'' in Table~\ref{tab:scs_msdialog}).
In particular, most of the utterances produced by the Assistant in MSDialog-Intent contain a request for feedback, which was not annotated by the crowd workers.

We observed that the annotation schemas could not be directly mapped in some cases (see different examples of ``Intent clarification'' in Table~\ref{tab:scs_msdialog}, which may be Initiative as well as NonInitiative).
% Transcriptions of spoken dialogues do not reflect intonation (see examples of ``Asks to repeat'' in Table~\ref{tab:scs_msdialog} that are interpreted as NonInitiative without a question mark);
Designing a guideline for the manual annotation task is challenging and error-prone (see questions and requests highlighted in Table~\ref{tab:scs_msdialog} but ignored by the human annotators).
Our automated utterance classifier is based on shallow features and designed to discriminate only between sentence types (interrogative and imperative versus declarative sentences).
% While the results of automated classifier 
In this evaluation, we showed that the results of our classifier not only correlate with the human judgement of utterance type but can be also used to infer missing labels.
% , proves to be a reliable indicator of initiative in dialogue that can be employed in the dataset analysis for comparing dialogue datasets on the large scale.

% The dialogue flow diagrams produced from manual and automatic QRFA labels illustrate discrepancies between the classes and their effect on the dialogue analysis results (see Figure~\ref{fig:scs}.

% By manually examining a random sample of errors produced by the classifier we identify two main sources of errors: 
% Figure~\ref{fig:scs}

% Note, how some of the examples annotated as questions do not contain an explicit question (see ``Original Question'', ``Intent clarification'' and ``Asks to repeat'' in Table~\ref{tab:scs_msdialog}), while other utterances that contain explicit questions or requests were not annotated as such (see ``Further Details'', ``Potential Answer'',  ``SERP with modification + Interpretation'' in Table~\ref{tab:scs_msdialog}).

% \usepackage{multirow}
\begin{sidewaystable}
\centering
\mbox{}\vspace*{11.5cm}
\caption{Mapping between MSDialog-Intent and SCSdata annotations to the QRFA labels. The sample utterances are shortened due to their length. \textit{Cursive} indicates examples of misclassification, in which information requests were not manually annotated in the original datasets.}
\label{tab:scs_msdialog}
\begin{tabular}{ll c l}
\toprule
\bf Dataset & \bf Original label & \bf QRFA & \bf Sample utterance \\
\midrule
\multirow{12}{*}{\begin{tabular}[c]{@{}l@{}}MSDialog-\\ Intent\end{tabular}} & Original Question & Q & Hello, I have done a reset ... I get to the same error. What do I do next? \\
& Original Question & Q & I called ... Note that I never ... which I cant. \\
& Information Request & R & What is the model of the computer? Have you tried ...? \\
& Follow Up + Repeat Question & AR & Hey there, just did a factory reset... Did you find a solution in the end ...? \\
& Further Details & F & no updates available I know of-\textit{please send me a list of any} ...\\
& Greetings/Gratitude + Positive Feedback & A & Hi... Thank you for posting back with the result. Glad to know the issue \\
& & & resolved. \textit{Feel free to post us if you need any assistance} ...\\
& Potential Answer & A & Once that you've restarted your computer, we suggest that you ... \\
& & & \textit{We would like to know if there is an antivirus software installed on your} \\
& & & \textit{computer. We look forward for your response.} \\
\midrule
\multirow{13}{*}{SCSdata} & Initial information request & Q & In which countries... in which European countries do they grow cinnamon? \\
& Access source & Q &  Can you just look at the news dot com \\
& Intent clarification & Q & Oh I'm I'm looking to find out uhm what what jobs are being outsourced \\
& & & from the US specifically to India \\
& Intent clarification & Q & Yes that's that's twenty to twenty-four \\
& Asks to repeat & R & Passenger and \\
& Asks to repeat & R & Oh per person \\
& Asks to repeat & R & Sorry \\
& Feedback on what is happening & F & So I'll ask a second question \\
& SERP with modification + Interpretation & A & It just says a lot of comparing and uhm like there are some articles that \\
& & & start to talk about like uhm sort of plants and stuff \\
& SERP with modification + Interpretation & A & The next one is the impact ... \textit{are you wanting anything newer?} \\
\bottomrule
\end{tabular}
\end{sidewaystable}

\begin{figure}[h]
  \centering
  \includegraphics[width=.98\linewidth,fbox]{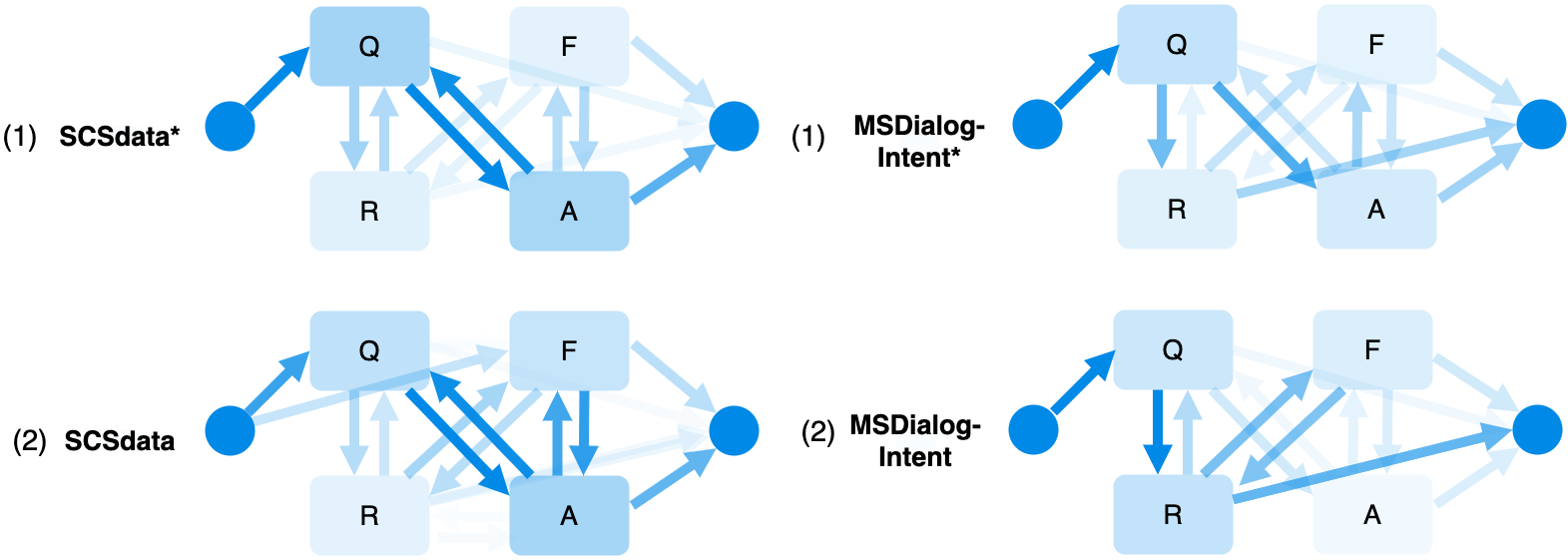}
  \caption{Dialogue flow diagrams with (1) manually and (2) automatically annotated utterance types.}
  \label{fig:scs}
\end{figure}

% results on the flow diagrams
Figure~\ref{fig:scs} shows the dialogue flow diagrams extracted from the SCSdata and MSDialog-Intent datasets:
\begin{enumerate}
	\item manually annotated by the dataset authors (SCSdata*) or crowdsourced (MSDialog-Intent*);
	\item automatically annotated with our utterance classifier (SCSdata and MSDialog-Intent). 
\end{enumerate}

\section{Results}
\label{sec:results}

% \todo{section overview}
In this section, we compare the dialogue datasets introduced in Section~\ref{sec:datasets} using the dialogue flow diagrams and asymmetry metrics.
% \todo{1) Dialogue Flow Analysis Results 2) Asymmetry Metrics Analysis Results}
We start with the dialogue flow analysis and use it to compare information-seeking dialogues with other dialogue types.

Dialogue flow diagrams provide a good overview of dialogue datasets but it is not easy to use them to compare, or to quantify similarities between, dialogues.
We show how the results obtained from the dialogue flow analysis can be further extended by adding other dimensions of mixed initiative offered by the asymmetry metrics.
The asymmetry metrics allow us to represent every dialogue as a vector and embed them into a common vector space.
We show that this approach makes it easier to compare dialogues along alternative dimensions and retrieve similar dialogues.

\subsection{Dialogue Flow Analysis Results}
\label{sec:dfresults}

\begin{figure}[h]
  \centering
  \includegraphics[width=0.95\linewidth,fbox]{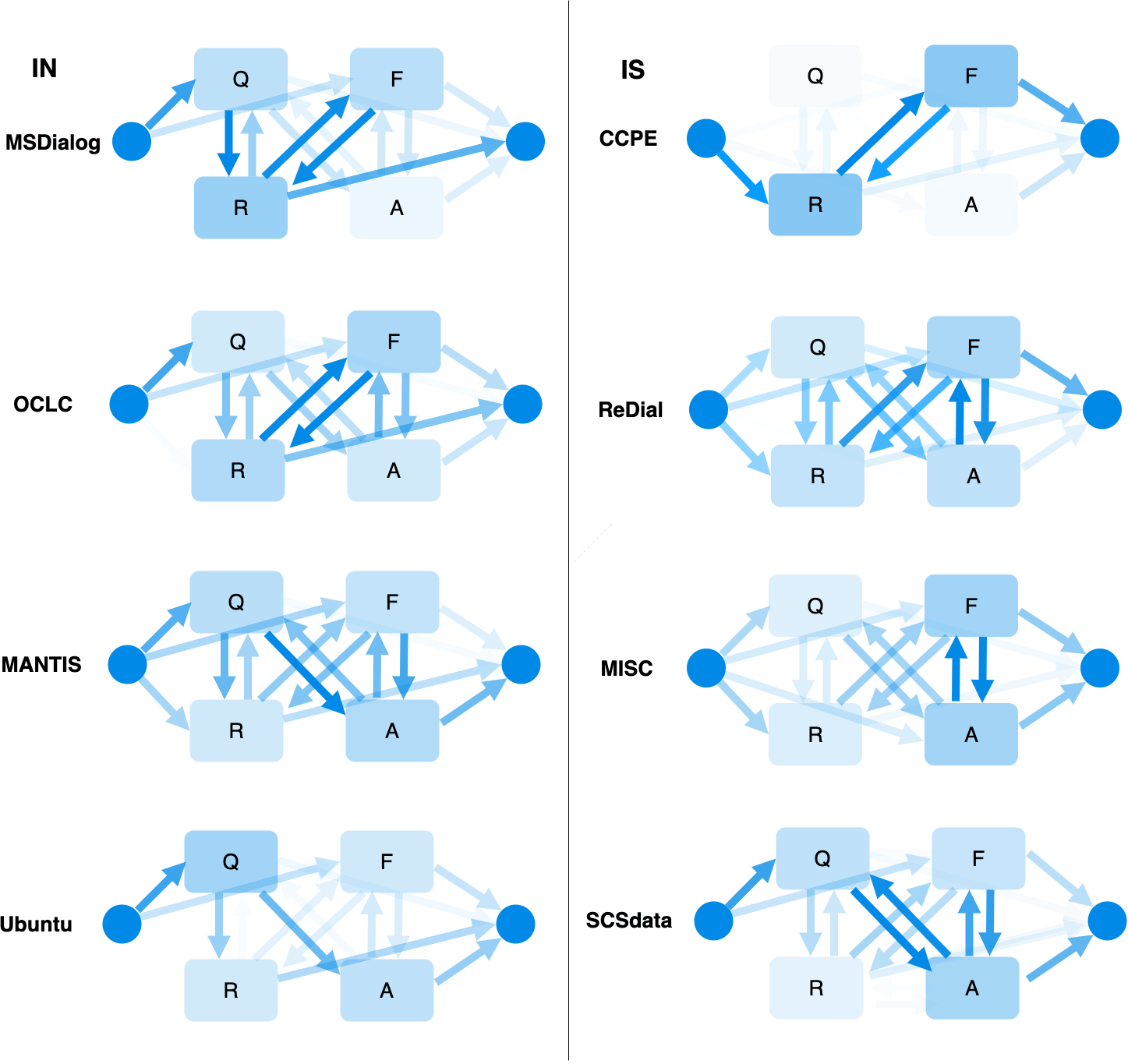}
  \caption{Dialogue flow in information-seeking dialogues collected in natural (IN) and simulated (IS) settings.}
  \label{fig:is}
\end{figure}

The dialogue flow diagrams produced for information-seeking dialogues are presented in Figure~\ref{fig:is}.
They are grouped into natural (IN, on the left) and simulated dialogues (IS, on the right).

The first thing to notice is that the diagrams vary a lot.
Therefore, we conclude that there are different subtypes of information-seeking dialogues according to the patterns of initiative.
% \mdr{which schema? and why not?}
This observation motivates us to reconsider the existing dialogue classification schema presented in Table~\ref{tab:datasets} and establish a new classification based on the patterns of mixed initiative.

\subsubsection{Search-Support dialogue classification}
\label{criterion}

Most of the dialogues in Figure~\ref{fig:is} demonstrate asymmetry of initiative towards one of the roles.
We refer to the dialogues in which the Seeker asks most of the questions and the Assistant provides most of the answers ($QA > RF$), as \textit{Search} dialogues.
For examples of Search dialogues in Figure~\ref{fig:is}, see MANTIS, Ubuntu, and SCSdata.

Search dialogues are contrasted with \textit{Support} dialogues, in which the Assistant plays a more active role by asking more questions and requesting information from the Seeker ($RF > QA$ ).
For examples of Support dialogues in Figure~\ref{fig:is}, see MSDialog, OCLC, CCPE and ReDial.
In CCPE almost all questions were asked by the Assistants since it is a preference elicitation dataset.
In comparison, ReDial has a smaller difference between the roles (82\% RF versus 56\% QA bigrams).
% , which is not so prominent in the diagram since most of the exchanges were recognised as NonInitiative (FA-AF).

% It is important to note that a Support dialogue does not become a Search dialogue by flipping the roles.
% All information-seeking dialogues (except for Ubuntu) were annotated with roles that indicate the dialogue participant with the primary information need.
% When the Assistant takes over the initiative by asking follow up questions, it indicates the secondary information need of the Assistant, who requires additional information in order to resolve the primary information need of the Seeker.
% This type of interaction is fundamentally different from a Search dialogue, in which the Seeker attempts to resolve his/her own information need be continuously asking follow up questions.

\subsubsection{Other dialogue types}

\begin{figure}[h]
  \centering
  \includegraphics[width=.98\textwidth,frame]{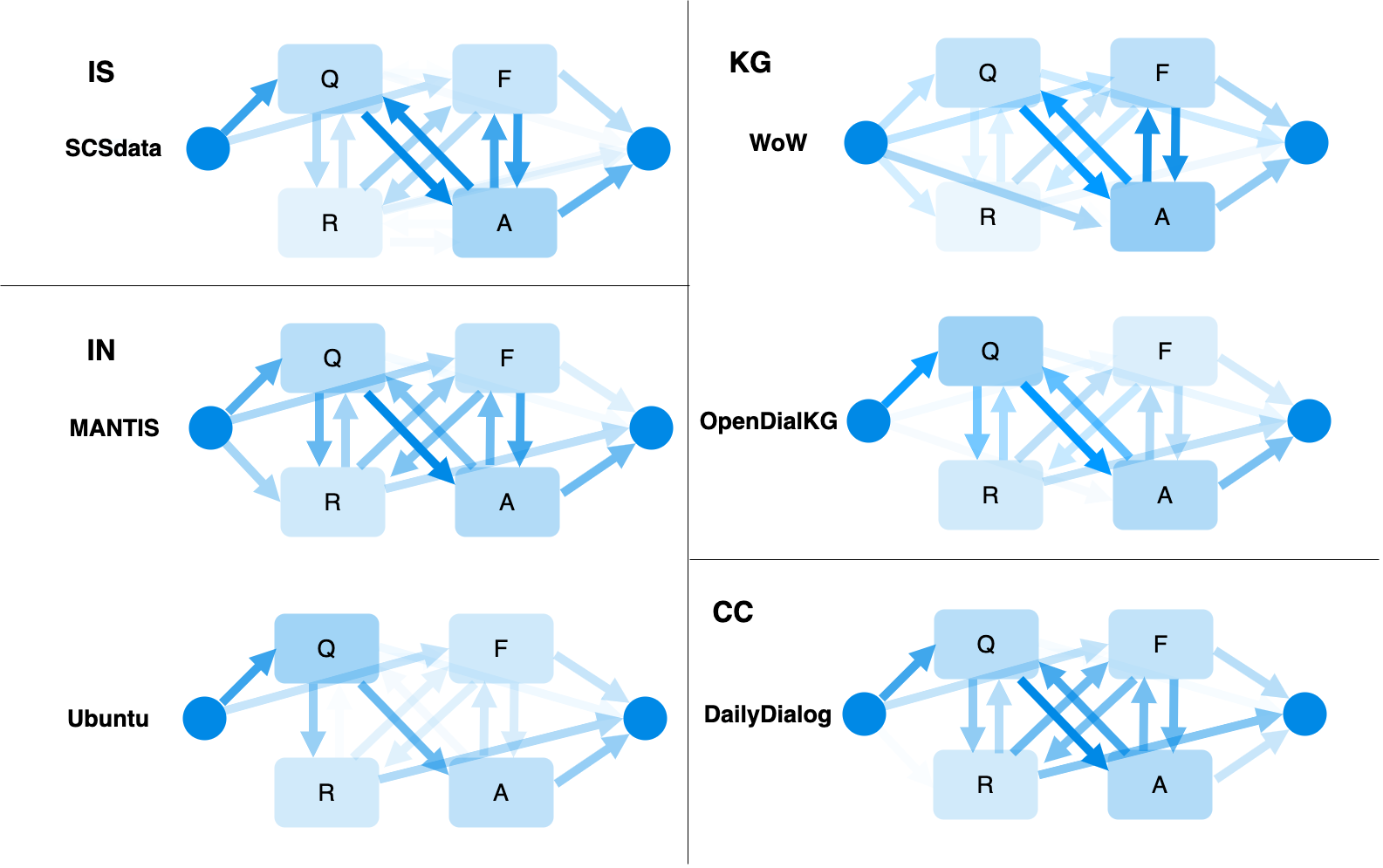}
  \caption{Search dialogues with initiative skewed towards the Seeker asking most of the questions ($QA>RF$).}
  \label{fig:search}
\end{figure}

For comparison we also produced dialogue flow diagrams for other dialogue types than information-seeking dialogues.
All dialogues are grouped by following the Search-Support classification criterion introduced in Section~\ref{criterion}:
\begin{align}
QA & > RF \rightarrow Search \\
QA & < RF \rightarrow Support
\end{align}
We found that \textit{Search} dialogues turn up very often among other dialogue types, including knowledge-grounded and chit-chat conversations (see Figure~\ref{fig:search}).
These dialogues exhibit similar interaction patterns as the ones observed in information-seeking dialogues extracted from on-line community discussions (MANTIS and Ubuntu).

\begin{figure}[h]
  \centering
  \includegraphics[width=.98\textwidth,fbox]{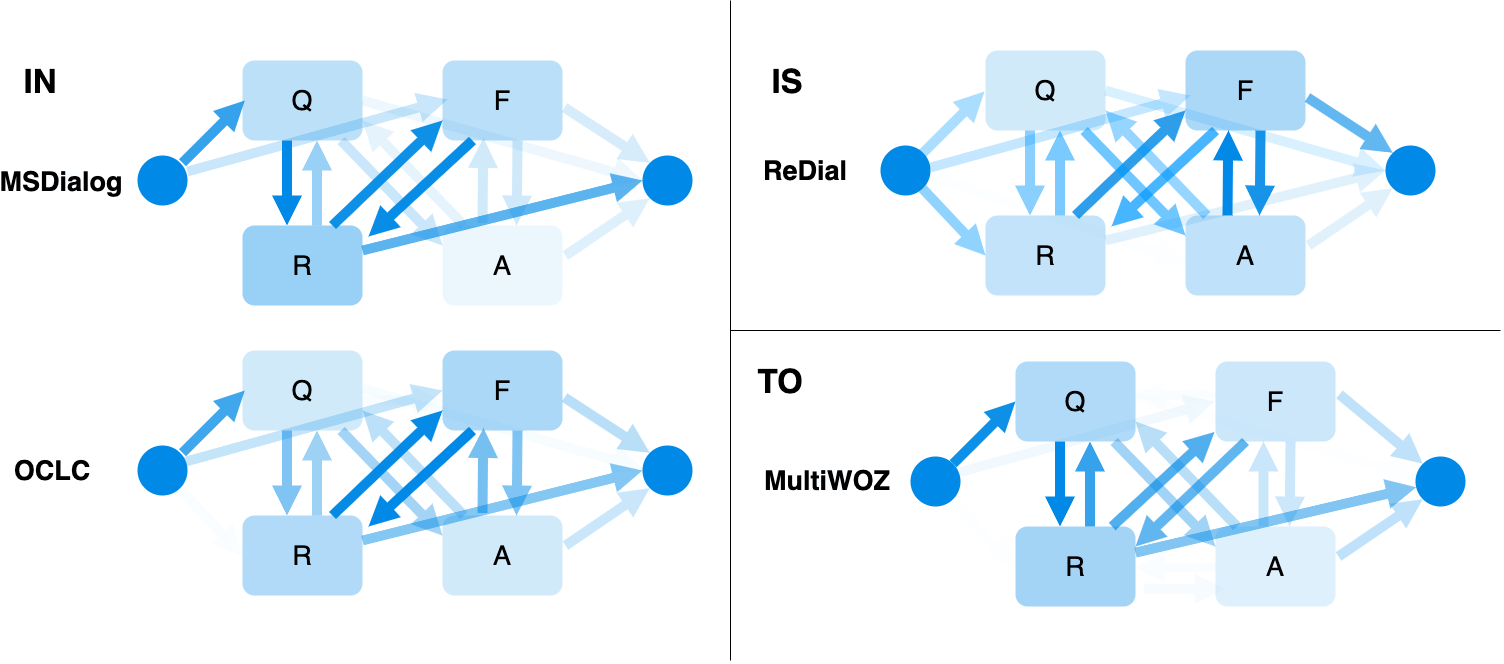}
  \caption{Support dialogues with initiative skewed towards the Assistant asking most of the questions (QA<RF)}
  \label{fig:support}
\end{figure}

Dialogues classified as Support (both information-seeking and non information-seeking) are shown in Figure~\ref{fig:support}.
Dialogues with this interaction pattern are produced either 
\begin{enumerate*}
\item by professional intermediaries in on-line forums and chat-rooms; 
\item in a conversational recommendation setting; or 
\item in a task-oriented dialogue setting.
\end{enumerate*}
It is also clear from Figure~\ref{fig:support} that none of the simulated datasets mirrors the patterns of initiative discovered in naturally occurring dialogues (MSDialog and OCLC).
ReDial has too much chit-chat (FA-AF) and MultiWOZ has too many questions by the Seeker (QR-RQ). 

\begin{figure}[h]
  \centering
  \includegraphics[width=.98\textwidth,fbox]{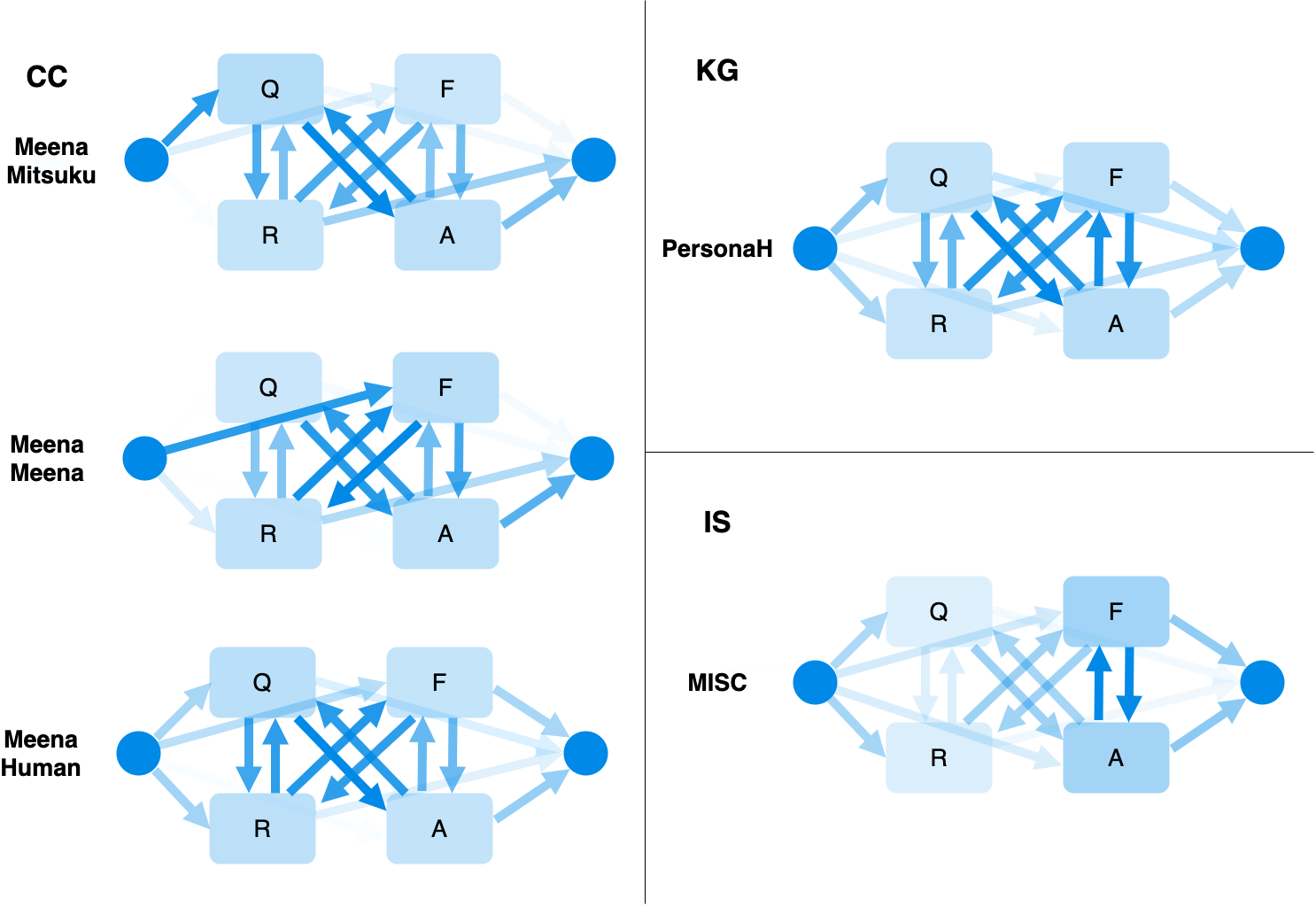}
  \caption{Information sharing dialogues appear symmetric due to equal distribution of initiative ($QA\approx RF$).}
  \label{fig:symmetry}
\end{figure}

Finally, in Figure~\ref{fig:symmetry} we show the dialogue flow diagrams for the datasets that do not fall into the previous two categories, i.e., in which $QA\approx RF$.
Notice how symmetry of speaker roles in social dialogues causes the diagrams to be symmetric as well.
In Meena and PersonaH datasets the speakers were instructed to have a casual chat without any specific roles assigned.
We refer to these dialogues as \emph{information-sharing} dialogues, in contrast with information-seeking dialogues.
% As seen in Figures~\ref{fig:support} and~\ref{fig:symmetry}, information-sharing dialogues are characterised by symmetry of initiative between the speakers, while the information-seeking dialogues are characterised by asymmetry of initiative either towards the Seeker (Search) or towards the Assistant (Support).

\revised{We established the differences between dialogues using the dialogue flow diagrams that cannot be explained by the initial dialogue classification provided in Table~\ref{tab:datasets}.
Figures~\ref{fig:search}-\ref{fig:symmetry} demonstrate that dialogues of different types often have similar structure and dialogues of the same type may exhibit very different structural patterns.
Therefore, we proposed to group similar dialogue datasets based on their structural patterns into three new dialogue types: Search, Support and Sharing.}

\subsection{Analysis Result of the Asymmetry Metrics}

\begin{figure}[!htb]
  \centering
  \includegraphics[width=.98\textwidth,frame]{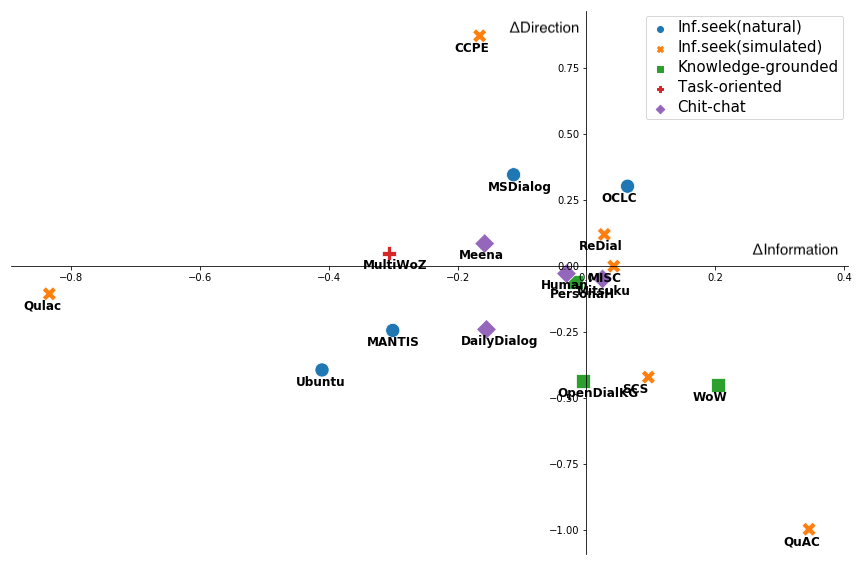}
   \caption{The scatter plot with dialogue datasets that shows the relation between $\Delta \mathit{Direction}$ and $\Delta \mathit{Information}$.}
  \label{fig:scatter_di}
\end{figure}

\begin{figure}[!htb]
  \centering
  \includegraphics[width=.98\textwidth,frame]{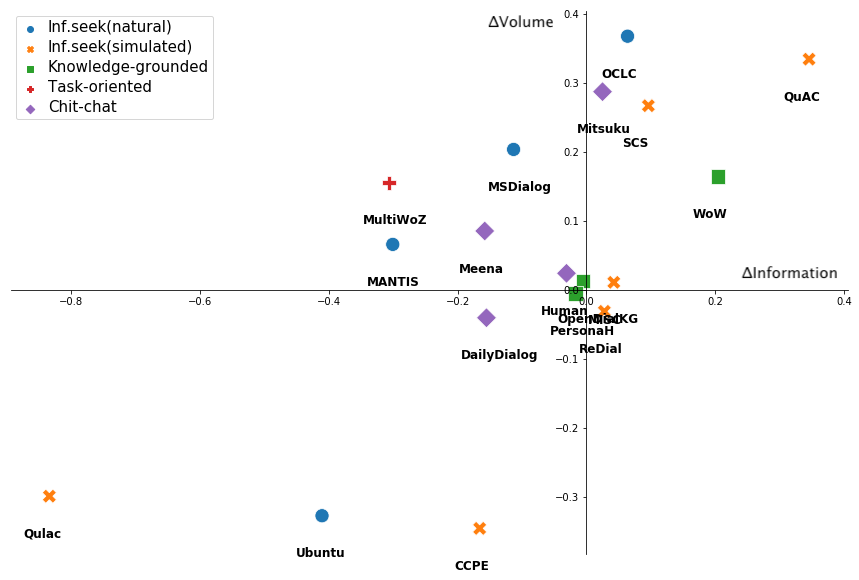}
%   \vspace{0.1cm}
  \caption{The scatter plot with dialogue datasets that shows the relation between $\Delta \mathit{Volume}$ and $\Delta \mathit{Information}$. }
  \label{fig:scatter_vi}
\end{figure}

\begin{figure}[!htb]
  \centering
  \includegraphics[width=.98\textwidth,frame]{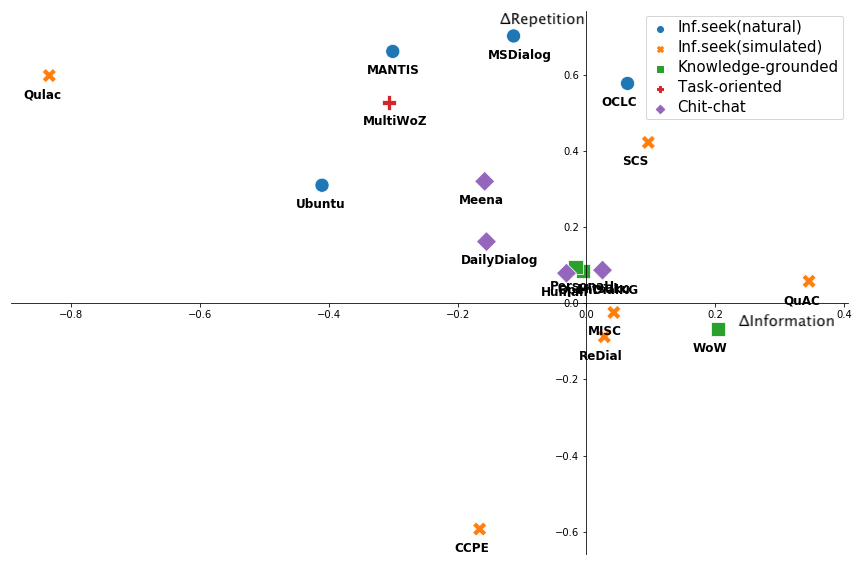}
%   \vspace{0.1cm}
  \caption{The scatter plot with dialogue datasets that shows the relation between $\Delta \mathit{Repetition}$ and $\Delta \mathit{Information}$. }
  \label{fig:scatter_ri}
\end{figure}

The results of embedding dialogue datasets using asymmetry metrics are displayed in Figures~\ref{fig:scatter_di}--\ref{fig:scatter_ri}.
In all plots, information sharing dialogues appear close to the origin of the coordinates, which represents a balance of initiative.
Information-seeking dialogues can be characterised by different types of asymmetries and are located away from the origin.

\begin{sidewaystable}
\centering
\mbox{}\vspace*{11.5cm}
\caption{Refined dataset types based on dimensions of mixed initiative identified via data flow diagrams and asymmetry metrics. The datasets are ordered by $\Delta \mathit{Direction}$. The negative values for all dimensions are highlighted in bold.}
\label{tab:asymmetries}
\begin{tabular}{lll rrrr}
\toprule
\textbf{Dataset} & \textbf{Type} & \textbf{Our Type} & \textbf{$\Delta$Volume} & \textbf{$\Delta$Direction} & \textbf{$\Delta$Information} & \textbf{$\Delta$Repetition} \\
\midrule
CCPE & Inf.seek(simulated) & Support & \textbf{-0.35} & 0.87 & \textbf{-0.17} & \textbf{-0.59} \\
MSDialog & Inf.seek(natural) & Support & 0.20 & 0.35 & \textbf{-0.11} & 0.70 \\
OCLC & Inf.seek(natural) & Support & 0.37 & 0.30 & 0.06 & 0.58 \\
ReDial & Inf.seek(simulated) & Support & \textbf{-0.03} & 0.12 & 0.03 & \textbf{-0.09} \\
Meena/Meena & Chit-chat & Sharing & 0.09 & 0.09 & \textbf{-0.16} & 0.32 \\
MultiWoZ & Task-oriented & Support & 0.16 & 0.05 & \textbf{-0.31} & 0.53 \\
MISC & Inf.seek(simulated) & Sharing & 0.01 & 0.00 & 0.04 & \textbf{-0.02} \\
Meena/Human & Chit-chat & Sharing & 0.02 & \textbf{-0.03} & \textbf{-0.03} & 0.08 \\
Meena/Mitsuku & Chit-chat & Sharing & 0.29 & \textbf{-0.05} & 0.02 & 0.09 \\
PersonaH & Knowledge-grounded & Sharing & \textbf{-0.01} & \textbf{-0.06} & \textbf{-0.02} & 0.10 \\
Qulac & Inf.seek(simulated) & Search & \textbf{-0.30} & \textbf{-0.10} & \textbf{-0.83} & 0.60 \\
DailyDialog & Chit-chat & Search & \textbf{-0.04} & \textbf{-0.24} & \textbf{-0.16} & 0.16 \\
MANTIS & Inf.seek(natural) & Search & 0.07 & \textbf{-0.24} & \textbf{-0.30} & 0.66 \\
Ubuntu & Inf.seek(natural) & Search & \textbf{-0.33} & \textbf{-0.39} & \textbf{-0.41} & 0.31 \\
SCSdata & Inf.seek(simulated) & Search & 0.27 & \textbf{-0.42} & 0.10 & 0.42 \\
OpenDialKG & Knowledge-grounded & Search & 0.01 & \textbf{-0.44} & 0.00 & 0.08 \\
WoW & Knowledge-grounded & Search & 0.16 & \textbf{-0.45} & 0.20 & \textbf{-0.07} \\
QuAC & Inf.seek(simulated) & Search & 0.33 & \textbf{-1.00} & 0.35 & 0.06 \\
\bottomrule
\end{tabular}
\end{sidewaystable}

% \todo{answers}
\revised{\textbf{Volume -- who talks more in a dialogue?}
In the majority of dialogue datasets that we consider in our analysis, the Assistant talks more than the Seeker.
There are only three datasets where the Seeker talks much more than the Assistant: CCPE, Qulac and Ubuntu ($\Delta \mathit{Volume} < - 0.25$).
% This result is not surprising for the CCPE and Qulac datasets since they were developed to collect samples of questions for search clarification and preference elicitation scenarios. 
The Ubuntu dataset is an outlier among the information-seeking dialogue datasets because it contains many samples where the questions posed by the Seekers were not answered.}

\revised{\textbf{Direction -- who requests information in a dialogue?}
The Seeker tends to ask more questions in the majority of datasets.
This does not hold for CCPE, MSDialog, OCLC, ReDial, Meena and MultiWoZ.
CCPE is the major outlier since most of the questions are asked by the Assistant to elicit user preferences.}

\revised{\textbf{Information -- who contributes to the dialogue topic? }
The scatter plot in Figure~\ref{fig:scatter_vi} shows that $\Delta \mathit{Volume}$ and $\Delta \mathit{Information}$ do not necessarily correlate.
In those cases, one of the speakers talks more but the dialogue topic is determined by the other speaker.
This is the case with the MSDialog and MANTIS datasets, which contain information-seeking dialogues extracted from forums.
More text is written by the Assistant but the repeated tokens are introduced by the Seeker.}
\revised{In the QuAC, SCS and WoW datasets, a different pattern is observed: the Assistant talks more and influences the conversation topic.}

% An important difference between these datasets and OCLC, however, remains. All three datasets are well below the y-axis in Figure~\ref{fig:scatter_di}. The higher $\Delta \mathit{Volume}$ and $\Delta \mathit{Information}$ in the simulated dialogues is achieved by the Assistant mainly answering questions. Since questions are usually shorter and contain less terms that have the potential to be repeated further on, this contributes to the skewed distribution of the Volume and Information metrics towards the Assistant in dialogues.
% In contrast, in the dialogues in the OCLC dataset the Assistant manages to dominate on all three asymmetry metrics, i.e., not only asking questions and providing answers but also actively contributing to the conversation topic, which is not observed in MSDialog and MANTIS.}

% There are several dialogue datasets for which this does not hold: MSDialog, Meena, MultWoZ, MANTIS.
% \todo{discuss which datasets are those and what it tells us}

\revised{\textbf{Repetition -- who follows up on the topic introduced by another participant?}
In the majority of dialogue datasets across all dialogue types, the Assistant tends to lead in the number of repetitions (see the points above the y-axis in Figure~\ref{fig:scatter_ri}).
However, this is mostly characteristic of the information-seeking and task-oriented dialogues.}

% \todo{types}

% \todo{discuss new metrics: information and volume}

% This observation reflects the situation in which the Assistant provides information, as measured by utterance length ($\Delta \mathit{Volume}$), but the Seeker does not make an apparent use of this information in the dialogue, i.e., does not follow up on the topics introduced by the Assistant, as measured by the repetitions ($\Delta \mathit{Information}$).

%   \smallskip

% \begin{figure}[h]
%   \centering
%   \includegraphics[width=.98\textwidth,frame]{figs/repetition-direction.png}
% %   \vspace{0.1cm}
%   \caption{The scatter plot with dialogue datasets that shows the relation between $\Delta \mathit{Direction}$ and $\Delta \mathit{Repetition}$. }
%   \label{fig:scatter_vr}
% \end{figure}

% \todo{OCLC}

\medskip\noindent%
\revised{To be able to simultaneously compare the datasets along all the dimensions of mixed initiative, we also present the results of the asymmetry metrics summarised into a single table (see Table~\ref{tab:asymmetries}).}
% The table shows that all four dimensions are independent. \mdr{independent of what? and how exactly is this shown?}}
\revised{All Support dialogues we identified in the dialogue flow analysis in Section~\ref{sec:dfresults} have a positive $\Delta \mathit{Direction}$, while they do not display a shared pattern for the other three metrics.
All Sharing dialogues have values close to zero, which we noticed already when looking at the individual plots.
The (single) task-oriented dialogue dataset, while having a $\Delta \mathit{Direction}$ value close to zero, differs from the Sharing dialogues on all other dimensions of initiative.
Finally, Search dialogues apart from the negative $\Delta \mathit{Direction}$ also predominantly have positive $\Delta \mathit{Repetition}$ (except for the WoW dataset).}

\revised{It is clear from Table~\ref{tab:asymmetries} that in Search dialogues the more questions come from the Seeker, i.e., $\Delta \mathit{Direction} \to -1$, the more information comes from the Assistant.
This dependency is much less pronounced in the Support dialogues.}
% \todo{comment on dialogue types we proposed and their characteristics across different dimensions}

\revised{Note that in contrast to all other datasets, the OCLC dataset, which contains virtual reference interviews, appears at the top of the upper-right quadrant in all three scatter plots.
The same result can be observed using Table~\ref{tab:asymmetries}: OCLC is the only dataset that has positive values across all four asymmetry metrics.
It shows that professional intermediaries play a very active role on all dimensions of initiative: asking questions, providing information, and introducing new subtopics that engage the Seeker.}

\section{Discussion}
\label{sec:discussion}

\revised{We started our analysis by introducing the dialogue datasets in Section~\ref{sec:datasets}.
Our description of their sources and the purpose for which they were collected, gave rise to initial hypotheses about their content and the dialogue types they represent.
We then proceeded to challenge these assumptions by systematically analysing and comparing structural patterns across dialogues.
% . who is asking questions who is providing information}
In the following, we summarise and discuss what we learned from this analysis with respect to (1) dialogue datasets for conversational search, (2) dialogue types and (3) relations between the dialogue analysis approaches we proposed.}

% \todo{give high-level overview of the subsections to come in this section and relations between them}

% \todo{contrast results from 2 approaches to initiative analysis and draw implications}

% conclusion
% The results of our analysis shows that none of the datasets collected to inform the design of conversational search systems exhibits the interaction patterns characteristic of an interview with a professional intermediary.
% or used for other tasks associated with conversation

\subsection{Conversational Search Datasets}
\label{sec:CSconcls}

\revised{Obtaining samples that resemble the target interaction style of an information-seeking conversation is vital for the development and evaluation of conversational search systems.}
The results of our analysis show that the dialogues produced to inform the design of a conversational search system, such as SCSdata and MISC, do not reflect the interaction patterns observed in the transcripts of virtual reference interviews (OCLC).
\revised{The differences observed are related to the distribution of mixed initiative between the dialogue participants. The Assistant is more passive in the simulated information-seeking dialogues, talking less and asking fewer questions (see Figures~\ref{fig:is}-\ref{fig:scatter_ri}).}

The results of our analysis indicate that the professional intermediaries in the OCLC dataset are more proactive than crowd workers and community experts, when measured on all dimensions of initiative.
They lead the conversation not only by writing long responses ($\Delta \mathit{Volume} > 0$) and asking many follow-up questions ($\Delta \mathit{Direction} > 0$) but also by actively steering the topic of a conversation ($\Delta \mathit{Information} > 0$).
These criteria should be considered in the evaluation of conversational search systems as well.

% \todo{a) the differences in the datasets;}

% \todo{b) what is “representative dialogue”; and, c) why it’s important to obtain representative dialogues.}
Our observations highlight that (1)~the setup of the data collection tasks is crucial for obtaining representative dialogues; and (2)~data analysis is crucial for verifying that the dialogues are representative.
% \todo{MISC as example}
For example, the MISC dataset was designed to inform the conversational search task.
However, we identified MISC as an information-sharing dialogue due to the symmetry of the speaker roles.

This finding is also in line with the observations reported by \citet{trippas2020towards}, who concluded that MISC contains chit-chat and negotiations by analysing these dialogues using manual annotations.
\revised{They suggest that the reason for this is the task setup.
The participants were not explicitly instructed on how they should interact.
The human intermediaries were provided with access to the information source but they were not instructed on how to conduct the interview.
In contrast, professional librarians and other domain experts, such as call-center personnel, receive specialised training on how to efficiently identify relevant aspects of an information need, assist and guide the Seeker during the interview.}

\revised{Our empirical evaluation demonstrates that the simulated information-seeking dialogue datasets, which were examined in our analysis, are not adequate for studying patterns of mixed initiative in dialogue.
We also showed that it is possible to perform the analysis of mixed initiative in dialogues automatically by comparing interaction patterns across dialogues.
This allowed us to scale such an analysis to thousands of dialogue transcripts sourced from a dozen of publicly available datasets.
The results show that neither of the publicly available datasets, which we considered in our analysis, matches the patterns extracted from the OCLC transcripts (see OCLC as an outlier in the scatter plots in Figures~\ref{fig:scatter_di}-\ref{fig:scatter_ri}).}

\revised{To move forward, the community should ensure that the conversational search systems that we design are modeled after professional intermediaries rather than the ad hoc strategies of non-expert volunteers.
As empirically shown in our analysis, the communication strategies of expert librarians differ considerably from the ones employed by the intermediaries in the simulated scenarios.}
% \todo{our role here}

% \todo{expand: limitations asr}
\revised{It is important to note also that MISC is the only dataset in our analysis that contains dialogue transcripts produced by an automated speech recogniser.
This likely lead to errors propagating from the utterance segmentation and classification steps into the analysis results.
For example, we observed that MISC contains very few questions, also in comparison to other social datasets in Figure~\ref{fig:symmetry}.
Nevertheless, our findings were confirmed by the results of the manual analysis performed independently by \citet{trippas2020towards}.
However, it is clear that to leverage data from spoken conversations our analysis techniques should be further adapted.
Currently, there is a lack of spoken information-seeking dialogues annotated with utterance labels, which are required for training and evaluation of utterance classifiers.}
% \todo{argue for future work}

\subsection{Dialogue Types}
\label{sec:DTconcls}

\revised{Careful collection and annotation of training data is crucial for developing successful machine learning models~\cite{roh2019survey}.
Therefore, it is also of a great importance to systematically analyse and correctly classify dialogues prior to using them for model training and evaluation.
This will allow us to avoid propagating biases towards undesirable patterns of interaction and dialogue characteristics in purely data-driven approaches, such as end-to-end dialogue models.}

% \mdr{huh:} In particular, the standard dialogue types cannot be reconstructed using the patterns of mixed initiative we considered here.

% \revised{Interestingly, we did not observe structural differences between the information-seeking dialogues collected to inform the conversational search task and the dialogues for other tasks, such as task-oriented, knowledge-grounded and chit-chat dialogues.
% On the contrary, we found that information-seeking dialogues vary a lot.
% The dialogue analysis techniques we proposed allowed us to distinguish different types of information-seeking dialogues that bear structural similarities to other dialogue types.}
% \todo{more concretely which similarities to which dialogue types}

% \todo{There are no such properties uncovered by the methods used, that can distinguish between data sets which have been identified as consisting of information-seeking dialogues and other dialogue types. One the other hand, the techniques could distinguish between different type of dialogues on some dimensions.}
% We discovered different types of information-seeking dialogues that bear structural similarities with other dialogue types, such as task-oriented, knowledge-grounded and chit-chat dialogues.

\revised{We uncovered important structural differences between the dialogue datasets that are generally considered to be of the same type.
For example, the information-seeking dialogues that were collected in different settings may have very different structure of mixed initiative (see Figure~\ref{fig:is}).}

\revised{Our results highlight that a common belief about a characteristic of a dialogue dataset can be incorrect.
For example, DailyDialog is a popular dataset for training data-driven dialogue models and considered to contain chit-chat dialogues~\cite{bao2019plato,sinha2020maude}.\footnote{\url{https://parl.ai/docs/tasks.html}}
These dialogues were collected from books for English learners and contain dialogues frequent in everyday situations, such as shopping or a job interview.
In our analysis, we discovered that this dataset differs from other chit-chat datasets.
This led us to discover that DailyDialog contains information-seeking dialogues, which also can be used to inform the conversational search task.
Those results emphasise that the content of dialogue datasets is often poorly understood.}

\revised{Such misconceptions about dialogue types may lead to wrong conclusions and misinterpretations of the experimental results.
For example, \citet{sinha2020maude} assumed that DailyDialog is a chit-chat dataset, which led them to conclude that their machine learning model ``captures the commonalities of chit-chat style dialogue'' without investigating what the commonalities of chit-chat style dialogues really are.}

\revised{Our results also indicate that information-seeking dialogues have similar dialogue flows as in task-oriented, grounded and chit-chat dialogues (see Figures~\ref{fig:search}-\ref{fig:symmetry}).
In most cases, apart from chit-chat, the initiative is skewed towards one of the dialogue participants.}
% These dialogue types are also not separable using more dimensions of initiative in Figures~\ref{fig:scatter_di}-\ref{fig:scatter_ri}.}

% \revised{We conclude that, when abstracting from the details of the system architecture and focusing on the system output alone, there is no conceptual difference between the dialogue tasks.}
% \revised{This finding implies that dialogues can be reused across tasks.
% In particular, the datasets of knowledge-grounded and task-oriented dialogues can be used to inform the design of conversational search systems since their dialogue structure is very similar.}

\revised{Overall, our results confirm that the current approach to classifying dialogues (see Table~\ref{tab:datasets}) based on the task and the system architecture does not adequately reflect the dialogue properties, in particular the patterns of mixed initiative that to a large extent characterise the type of a dialogue interaction (see Table~\ref{tab:asymmetries}).
Therefore, on the basis of our analysis, we introduced a new dialogue typology which reflects the communication strategies that are common across different dialogue datasets by measuring the degree of mixed initiative: Search, Sharing and Support (see Figures~\ref{fig:search}--\ref{fig:symmetry}). While this classification schema is relatively simple, we show that it can be further enhanced by considering additional dimensions of initiative (see Figures~\ref{fig:scatter_di}--\ref{fig:scatter_ri}, Table~\ref{tab:asymmetries}.)}

\subsection{Comparison of the Dialogue Analysis Approaches}
\label{sec:DAconcls}

\revised{The approaches to dialogue analysis we employed here shed light on the differences and similarities in patterns of mixed initiative.
Dialogue flow diagrams provide a convenient overview of the transition frequencies between dialogue turns of different type.
Their main benefit is that they provide a compact but explainable summary of the mixed initiative distribution.
The important drawbacks of the dialogue flow diagrams are that they are difficult to compare and only reflect a single dimension of mixed initiative, namely, the utterance type.
% twofold: (1) the diagrams are hard to compare since they rely on many components, such as unigram and bigram distributions;
We overcome these limitations by introducing asymmetry metrics.
In addition to quantifying the difference in utterance types, asymmetry metrics also reflect other dimensions of initiative: the distribution of utterance lengths and term repetitions.}

% \todo{asymmetry metrics}
\revised{The asymmetry metrics provide for sensible and meaningful dimensions that allow us to represent each dataset as a point in a vector space.
For example, the scatter plot in Figure~\ref{fig:scatter_di} was produced by using $\Delta \mathit{Information}$ and $\Delta \mathit{Direction}$ as x- and y-coordinates for each of the datasets.
This approach allow us to conveniently compare datasets in a vector space and identify similar datasets using standard metrics, such as euclidean or cosine distance.
This embedding approach also allows for results to be interpretable since we use the asymmetry metrics as dimensions, which reflect certain dialogue properties that are explicitly measured.}

% \todo{connection}
\revised{Importantly, the $\Delta \mathit{Direction}$ metric agrees with the results of our dialogue flow analysis.
This is evident from the plot in Figure~\ref{fig:scatter_di}, where Search and Support dialogues are located on opposite sides of the x-axis.
More specifically, all the datasets presented in Figure~\ref{fig:support}, i.e., MSDialog, OCLC, ReDial and MultiWoZ, have $\Delta \mathit{Direction} > 0$, while all datasets in Figure~\ref{fig:search}, i.e., SCSdata, MANTIS, Ubuntu, WoW, OpenDialKG and DailyDialog, have $\Delta \mathit{Direction} < 0$.}

\revised{Both dialogue flow diagrams and asymmetry metrics can be used in a combination to successfully leverage their advantages and alleviate their drawbacks. For example, when designing a repository for dialogue data the asymmetry metrics can provide facets to enable search and browsing interface, while the dialogue flow diagrams can serve as a summary of the dataset content.}

\section{Conclusion}
\label{sec:consclusion}
% what we did
\revised{We introduced a dialogue representation approach and two dialogue analysis approaches, that is, dialogue flow diagrams and asymmetry metrics, that allow one to compare the structure of dialogues from different datasets.
We applied both approaches to conduct a large-scale analysis of dialogue transcripts specifically focusing on the patterns of mixed initiative to distinguish information-seeking dialogues from other dialogue types.
The results of our analysis including the dialogue fingerprints\footnote{\url{https://uvaauas.figshare.com/articles/dataset/dialogue_fingerprints/13356350}} as well as the scripts required to reproduce them\footnote{\url{https://github.com/svakulenk0/conversation_shape}} are publicly available to encourage future work in this direction.}

\subsection{Lessons Learned}

Based on our analysis of the results and their discussion in the previous sections, we come back to answer the main research questions that were introduced in Section~\ref{sec:introduction}.

% \todo{Make sure to refer back to the discussion section (“as we argued in Section~6, …”).}
% \todo{conclude by answering RQs}

\begin{enumerate}
    \item[\textbf{RQ1}]What are the structural properties of information-seeking dialogues that differentiate them from other dialogue types?
\end{enumerate}

% what did we find? -> similarity of dialogue types
\noindent%
\revised{Interestingly, we did not observe structural differences between  information-seeking dialogues collected to inform the conversational search task and dialogues for other tasks, such as task-oriented, knowledge-grounded and chit-chat dialogues.
On the contrary, we found that information-seeking dialogues vary a lot. We discovered several types of information-seeking dialogues and showed that they bear structural similarities to other dialogues.}

\revised{On the basis of our analysis, we introduced a new dialogue typology that reflects the communication strategies that are common across different dialogue datasets: Search, Sharing and Support.
We believe that conversational search systems should move from Search towards (information) Sharing and Support by taking over more initiative in a conversation, as exhibited in virtual reference interviews with professional librarians.}

\begin{enumerate}
    \item[\textbf{RQ2}] Which datasets contain dialogues similar to virtual reference interviews with a professional librarian?
\end{enumerate}

% what did we find? -> missing dataset 
% % \todo{what characteristics of the other information-seeking dialogue datasets make them unsuitable as models for studying interaction, especially as compared to characteristics of the OCLC dataset.}
\noindent%
None of the dialogue datasets mirrors the patterns of mixed initiative in virtual reference interviews \revised{from the OCLC dataset.}
\revised{Thereby, we showed that the dialogues that were collected in a laboratory environment between volunteers using simulated information needs do not represent the dialogue type that naturally occurs between an expert intermediary and a seeker with a genuine information need.
Non-expert intermediaries write less and ask less questions than professional librarians.
This finding implies that the data collection procedures that the community has designed to inform the conversational search task requires adjustment and better quality control.}

% a more general statement on how good/convincing our current collection of datasets is
\revised{Thus, the results of our analysis provide evidence for the claim that existing datasets collected to inform the conversational search task (MISC, SCSdata, etc.) are not suitable for studying and designing mixed initiative systems, as we argued in Section~\ref{sec:CSconcls}.
The community should focus on more realistic datasets, such as OCLC, to better understand the patterns of initiative from interactions between skilled interviewers/librarians and information seekers.}

% lessons on our methodology
\revised{Overall, we showed that dialogue flow diagrams can provide a convenient overview for a dialogue dataset and they are easy to interpret, while asymmetry metrics allow us to conveniently compare dialogue datasets across several dimensions of mixed initiative simultaneously.
Both approaches enabled us to discover frequent patterns that provide valuable insights as to the important characteristics and qualities of the dialogue datasets currently available to the research community.
We consider this to constitute an important milestone towards establishing a mechanism for continuous refinement of our understanding of information-seeking dialogue interactions based on the growing volume of empirical data, while also evaluating the quality of the dialogue data being used for model development.}

% \todo{re: Our goal is to establish a mechanism that can enable us to continuously refine the theories of conversational search based on the growing volume of empirical data. how far did we get?}

\subsection{Future Work}
% \todo{limitation: classification accuracy}

The results of our large-scale dialogue analysis are based on automatic utterance classification and term-based repetition patterns.
Both of these approaches have their limitations and the errors introduced by them affect the results of dialogue analysis.
% \todo{expand}
% \paragraph{Limitations of our dialogue analysis approach.}
\revised{The metrics we introduced for measuring initiative in dialogue are relatively simple and crude.
We are likely to underestimate the number of questions by both participants due to errors in classification and the number of repetitions since the lexical matches capture only explicit references to the previous context overlooking more subtle semantic relations between the utterances.}
% \todo{is not able to capture more subtle signals e.g. co-reference (repetition) it is still hard for dialogue and simple anaphora counts can be misleading. We also experimented with an off-the-shelf co-reference resolution model instead but were not satisfied with the results.}
% \todo{but we show that still able to obtain reliable results. how do we know our results are reliable? validity evaluation?}
Nevertheless, our results demonstrate the value of automatic data analysis approaches that have implications on the data collection, interaction design and evaluation.
Future work should focus on the evaluation and improvement of the representational power of the dialogue analysis approaches.

% future work -> interface for dialogue dataset search

% Future work should focus on the evaluation and improvement of the representational power of the dialogue analysis approaches.
% While the approaches we proposed are not without limitations, our goal was to pave the way towards a more systematic analysis of dialogue datasets.
% This functionality can be integrated with a conversational data repository that will allow to enhance browsing and search for dialogue data that can be used for training and evaluation of dialogue systems.
% \todo{expand}

The interaction patterns we discovered can also be used to enhance browsing and search interfaces providing access to a repository of dialogue datasets.
Modeling structural properties of dialogues based on their content rather than merely their metadata description will help to retrieve samples that can be used for training and evaluation of dialogue models.
% \todo{expand}
% As was shown in our analysis, our expectation of the dialogue properties may 
% \todo{expand}

% \section{Model Diagnostics}
% \label{sec:model_diagnostics}

% \todo{Our approach can be also used to evaluate dialogue models}
% % We showed that our approach helps to detect defficiencies and strategies used by dialogue models.
% This approach can be useful for model selection and training as well.
% \todo{expand}

\begin{acks}
We thank our anonymous reviewers for extensive comments and suggestions that helped us to improve the paper.

This research was inspired by the discussions held at the Dagstuhl Seminar 19461 on Conversational Search.
We thank the organisers and participants of the seminar, and, especially, Filip Radlinski and Nicholas Belkin for triggering the idea for this study.
We are especially grateful to Marie Radford, Lynn S. Connaway and Andrew K. Pace for providing us with the dataset of virtual reference interviews from OCLC.

This research was supported by
the Netherlands Organisation for Scientific Research (NWO),
Innovational Research Incentives Scheme Vidi (016.Vidi.189.039) and
Smart Culture - Big Data/Digital Humanities (314-99-301),
the H2020-EU.3.4. - SOCIETAL CHALLENGES - Smart, Green And Integrated Transport (814961),
the Google Faculty Research Awards program,
and the Hybrid Intelligence Center, 
a 10-year program
funded by the Dutch Ministry of Education, Culture and Science through 
the Netherlands
Organisation for Scientific Research, 
\url{https://hybrid-intelligence-centre.nl}.

All content represents the opinion of the authors, which is not necessarily shared or endorsed by their respective employers and/or sponsors.
\end{acks}

\bibliographystyle{ACM-Reference-Format}
\bibliography{refs}

\end{document}